\documentclass[11pt]{article}
\pdfoutput=1
\usepackage[utf8]{inputenc}
\usepackage[left=20mm,right=20mm,top=25mm,bottom=25mm]{geometry}
\usepackage[T1]{fontenc}

\usepackage{graphicx} 
\graphicspath{{./figs}}

\usepackage{siunitx}
\usepackage{subcaption}
	
\usepackage{pgfplots}
\pgfplotsset{compat=newest} 
\pgfplotsset{plot coordinates/math parser=false} 

\usepgfplotslibrary{external} 
\tikzsetexternalprefix{TikzFig/}
\pgfkeys{/pgf/number format/.cd,1000 sep={}} 

\pgfplotsset{every axis/.append style={
     legend style={nodes={font=\tiny,scale=1.2}} 
  		},
		every tick label/.append style={scale=0.8}, 
   }
\pgfplotsset{
  every axis plot/.append style={line width=0.55pt},
  every axis plot post/.append style={
    every mark/.append style={line width=0.6pt}
  }
}

\usetikzlibrary{backgrounds} 
\pgfdeclarelayer{background}
\pgfdeclarelayer{foreground}
\pgfsetlayers{background,main,foreground} 

\newlength\fheight
\newlength\fwidth
\newlength\figureheightbig
\newlength\figurewidthbig

\usepackage{outlines}

\usepackage{booktabs} 
\usepackage{amsmath} 

\usepackage{algorithm} 
\usepackage{algpseudocode} 

\usepackage{afterpage}
\usepackage{tabularx}
\usepackage{csquotes}

\usepackage{hyperref} 
\hypersetup{
    colorlinks=true,
    linkcolor=tud2a,
    filecolor=magenta,      
    urlcolor=tud2a,
    citecolor=tud2a,
    pdfpagemode=FullScreen,
    }

 \usepackage[affil-it]{authblk}
\usepackage[maxbibnames=10, maxcitenames=3, giveninits=true,
    natbib=true, bibencoding=utf8, isbn=false, url=false, backend=bibtex, style=authoryear-comp]{biblatex} 
\AtEveryBibitem{\clearfield{month}}
\AtEveryBibitem{\clearfield{day}}
\AtEveryBibitem{\clearlist{language}}
 
\definecolor{tud1a}{HTML}{5D85C3}
\definecolor{tud2atrue}{HTML}{009CDA}
\definecolor{tud2a}{rgb}{0.00000,0.44700,0.74100}%
\definecolor{tud3a}{HTML}{50B695}
\definecolor{tud3ai}{HTML}{009a67} 
\definecolor{tud4a}{HTML}{AFCC50}
\definecolor{tud5a}{HTML}{DDDF48}
\definecolor{tud6a}{HTML}{FFE05C}
\definecolor{tud6b}{HTML}{FDCA00}
\definecolor{tud7b}{HTML}{F5A300} 
\definecolor{tud8a}{HTML}{EE7A34}
\definecolor{tud8b}{HTML}{EC6500}
\definecolor{tud9a}{HTML}{E9503E}
\definecolor{tud11b}{HTML}{721085}
\definecolor{tud1b}{HTML}{005AA9} 
\definecolor{tud3b}{HTML}{009D81}
\definecolor{tud4b}{HTML}{99C000}
\definecolor{tud4c}{HTML}{7FAB16}
\definecolor{tud4d}{HTML}{6A8B22}
\definecolor{tud9b}{HTML}{E6001A}
\definecolor{tud9c}{HTML}{B90F22} 
\definecolor{tud10a}{HTML}{A60084} 
\definecolor{sourcecode}{HTML}{E9E9E9}
\definecolor{mygrey}{HTML}{C0C0C0}

\DeclareMathOperator*{\mymax}{max}
\DeclareMathOperator*{\mymin}{min}

\newcommand{\vect}[1]{\boldsymbol{#1}}

\newcommand{\overbar}[1]{\mkern 1.5mu\overline{\mkern-1.5mu#1\mkern-1.5mu}\mkern 1.5mu}

\newcommand{\Arch}{\operatorname{\mathit{A\kern-.06em r}}} 
\newcommand{\Biot}{\operatorname{\mathit{B\kern-.06em i}}} 
\newcommand{\Cauc}{\operatorname{\mathit{C\kern-.07em a}}} 
\newcommand{\Damk}{\operatorname{\mathit{D\kern-.06em a}}} 
\newcommand{\Eule}{\operatorname{\mathit{E\kern-.03em u}}} 
\newcommand{\Four}{\operatorname{\mathit{F\kern-.10em o}}} 
\newcommand{\Frou}{\operatorname{\mathit{F\kern-.07em r}}} 
\newcommand{\Gras}{\operatorname{\mathit{G\kern-.05em r}}} 
\newcommand{\Karl}{\operatorname{\mathit{K\kern-.11em a}}} 
\newcommand{\Knud}{\operatorname{\mathit{K\kern-.11em n}}} 
\newcommand{\Lewi}{\operatorname{\mathit{L\kern-.05em e}}} 
\newcommand{\Mach}{\operatorname{\mathit{M\kern-.10em a}}} 
\newcommand{\Nuss}{\operatorname{\mathit{N\kern-.09em u}}} 
\newcommand{\Pecl}{\operatorname{\mathit{P\kern-.08em e}}} 
\newcommand{\Pran}{\operatorname{\mathit{P\kern-.03em r}}} 
\newcommand{\Rayl}{\operatorname{\mathit{R\kern-.04em a}}} 
\newcommand{\Reyn}{\operatorname{\mathit{R\kern-.04em e}}} 
\newcommand{\Rich}{\operatorname{\mathit{R\kern-.06em i}}} 
\newcommand{\Schm}{\operatorname{\mathit{S\kern-.07em c}}} 
\newcommand{\Sher}{\operatorname{\mathit{S\kern-.07em h}}} 
\newcommand{\Stro}{\operatorname{\mathit{S\kern-.07em r}}} 
\newcommand{\Webe}{\operatorname{\mathit{W\kern-.14em e}}} 

\title{
Physics-based Digital Twins for Autonomous Thermal Food Processing: Efficient, Non-intrusive Reduced-order Modeling
}

\author[1,*]{Maximilian~Kannapinn}
\author[2]{Minh~Khang~Pham}
\author[1]{Michael~Schäfer}
\affil[1]{\footnotesize Institute for Numerical Methods in Mechanical Engineering \& Graduate School of Computational Engineering, \protect \\
Department of Mechanical Engineering \& Centre for Computational Engineering,\protect \\ Technical University of Darmstadt, Dolivostr.~15, 64293 Darmstadt, Germany}
\affil[2]{\footnotesize Technical University of Darmstadt, Karolinenplatz  5, 64289, Darmstadt, Germany}
\affil[*]{\footnotesize Corresponding author, Email: \href{mailto:research@maxkann.de}{research@maxkann.de}}

\date{September 04, 2022}

\usepackage{amssymb}

\usepackage{lineno}

\begin{document}
\maketitle


%
%
%
 \par\noindent\rule{\textwidth}{0.4pt}           
\begin{abstract} \noindent
One possible way of making thermal processing controllable is to gather real-time information on the product's current state. Often, sensory equipment cannot capture all relevant information easily or at all. Digital Twins close this gap with virtual probes in real-time simulations, synchronized with the process. This paper proposes a physics-based, data-driven Digital Twin framework for autonomous food processing.
We suggest a lean Digital Twin concept that is executable at the device level, entailing minimal computational load, data storage, and sensor data requirements. 
This study focuses on a parsimonious experimental design for training non-intrusive reduced-order models (ROMs) of a thermal process. 
A correlation ($R=-0.76$) between a high standard deviation of the surface temperatures in the training data and a low root mean square error in ROM testing enables efficient selection of training data. The mean test root mean square error of the best ROM is less than 1 Kelvin ($ \SI{0.2}{\%}$ mean average percentage error) on representative test sets. Simulation speed-ups of $\operatorname{Sp} \approx \SI{1.8E4}{}$ allow on-device model predictive control.

 \paragraph*{Industrial relevance}
The proposed Digital Twin framework is designed to be applicable within the industry. Typically, non-intrusive reduced-order modeling is required as soon as the modeling of the process is performed in software, where root-level access to the solver is not provided, such as commercial simulation software. The data-driven training of the reduced-order model is achieved with only one data set, as correlations are utilized to predict the training success a priori.
 
\end{abstract}






\vspace*{0.5ex}
\par\noindent\rule{\textwidth}{0.4pt}\vspace*{2pt}
{\small
Accepted version of manuscript published in \emph{Innovative Food Science and Emerging Technologies}. \\
Date accepted: September 04, 2022. 
License: \href{https://creativecommons.org/licenses/by-nc-nd/4.0/legalcode}{CC BY-NC-ND 4.0}
}
\vspace*{-1.6mm}
\par\noindent\rule{\textwidth}{0.4pt}



\renewcommand{\sectionautorefname}{Sec.}
\renewcommand{\subsectionautorefname}{Sec.}
\renewcommand{\subsubsectionautorefname}{Sec.}
\newcommand{\algorithmautorefname}{Algorithm}
\widowpenalty10000
\clubpenalty10000

\renewcommand{\appendixname}{Appendix}

\newpage
\section{Introduction}

The global population is expected to exceed 10 billion people by 2050. The food industry thus faces the challenge of improving food supply sustainability, becoming more energy-efficient, and reducing food waste while additionally complying with customer demands for nutritious and safe food.
For years, scientific reports have demanded that humankind's carbon footprint be reduced to maintain global temperatures at a safe level.
A special report by the Intergovernmental Panel on Climate Change attributes \si{21}--\SI{37}{\%} of total net anthropogenic greenhouse gases to the food system, whereas \si{8}--\SI{10}{\%} can be attributed to food wastage~(\cite{mot_ipcc_climate2019}). 
The food system alone has a \SI{67}{\%} probability of endangering the \SI{2}{\degreeCelsius} greenhouse gas reduction target by 2050 if no change is undertaken~(\cite{Clark2020}). 

Nonetheless, food safety must remain the top priority when developing novel technologies to improve energy efficiency. \cite{mot_langsrud_cooking2020} disclosed deficits in judging the doneness of food, as approximately \SI{30}{\%} of food-borne illnesses can be related to eating undercooked poultry. Large quantities of safe, healthy, quality meals are required for community catering, such as in hospitals, schools, universities, diners, and canteens, or for crisis response. In these areas, predominantly less-trained personnel operate. In general, the food preparation sector has faced a shortage of skilled staff in recent years. Even before the coronavirus pandemic, among Germany's top six most unpopular jobs were chef and sous-chef~(\cite{mot_zeit2018}). 
Hence, adequate supervision and decision support are required in kitchens.
Novel ways of reducing energy consumption and food waste in the private, industrial, and public sectors while preserving or improving food quality are also needed.
Minor improvements can have a significant impact when applied at scale.
Autonomous processes with artificial intelligence are among the next disruptive technological trends, and Digital Twins might act as enablers~(\cite{Rosen2015}).
Before a brief overview of the state of the art of Digital Twins in food science and technology is provided, the following section summarizes the key features of Digital Twins.

\paragraph{Definition of Digital Twins}
 
 Michael Grieves originally derived the concept of Digital Twins in 2003 as the \enquote{mirrored spaces model} for his lectures at the University of Michigan, and years later, Vickers inspired the name Digital Twin (\cite{Grieves2017}). Approximately \SI{85}{\%} of the Digital Twin studies to date focused on product life-cycle management and \SI{11}{\%} focused on factory or manufacturing planning~(\cite{Lu2020}).
Various efforts have also been made to define and standardize the Digital Twin concept~(e.g.,~\cite{dt_Stark2019,dt_AIAA2020,dt_ISO23247}), and it has been identified as an active field of research~(\cite{dt_tao2019,Niederer2021}).
Common features that characterize a Digital Twin can be found in literature (see summary in \autoref{tab:DTcriteria}).
\begin{table*}[bhtp]
\scriptsize
\centering
\addtolength{\tabcolsep}{-0pt}
 \caption{Common features of the Digital Twin concept in literature.} 
\begin{tabularx}{\textwidth}{l l X} \toprule 
\textbf{Common Feature} & \textbf{Description}                                                                               		& \textbf{Sources}                                                                           \\
\midrule 
F1: Digital Mirror          & Consists of a simulation model / a virtual domain                                                 	& \cite{Grieves2017,dt_AIAA2020,Rasheed2020}                                \\
F2: Speed-up              & Must reflect the physical process in real-time                                          	&  \cite{Glaessgen2012,Rasheed2020,Lu2020}                                  \\%
F3: High-fidelity           & Must reflect the physical process at utmost accuracy                       & \cite{Glaessgen2012,Grieves2017,Rasheed2020}                      \\ 
F4: Symbiosis               & Synergies of physics-based and data-based modeling                                                 &  \cite{Verboven2020,Niederer2021,Rasheed2020}                       \\
F5: Bi-directionality       & Mutual data exchange of Digital and Physical Twin                                                                  		& \cite{Glaessgen2012,dt_AIAA2020,Niederer2021}                     \\
F6: Persistence             & Evolves during the product life-cycle and reflects changes                     & \cite{Grieves2017,dt_AIAA2020,Rasheed2020}                                              \\
F7: Robustness              & Reliable / accounts for uncertainty / validated                                                 		&  \cite{Niederer2021,Rasheed2020}                                                                \\
F8: Explorative             & Monitoring / automation / autonomy 		& \cite{Grieves2017,dt_AIAA2020,Niederer2021,Rosen2015} \\ %
\bottomrule 
 \label{tab:DTcriteria}
\end{tabularx}
\end{table*}
A Digital Twin can be defined as a virtual set of information (labeled F1 in \autoref{tab:DTcriteria}) replicating its physical counterpart at the utmost fidelity (F3). Digital and Physical Twins engage in bi-directional data exchange (F4, F5) in real time (F2). 
This definition clarifies that the concept of a Digital Twin is a much more sophisticated than that of a simulation model, although this is still a common misconception~(\cite{dt_tao2019,Lu2020}). 

Digital Twins symbolize the endeavor to make simulation and data science technologies available not only in the design phase (\enquote{what-if simulations}) but also in the operational phase (F6)~(\cite{Niederer2021}). In the latter phase, people and potentially autonomous processes might make better-informed decisions (F8)~(\cite{Rasheed2020}).
However, how to deploy Digital Twins at scale robustly (F7) is still an open research question as novel mathematical, numerical, and computational methods first have to be developed~(\cite{Niederer2021}).

 \paragraph{State of the Art} 
We have witnessed the advent of Digital Twins in the field of food science and technology in the last two years, which explains why the relevant literature is still scarce. 
A recent literature review by \cite{dt_henrichs2022} elucidates that few food science studies considered autonomy of processes. Only eight out of 84 studies focused on Digital Twin-enabled autonomy of processes, and of these eight, only two sources were peer-reviewed studies. \cite{dt_guo2018} monitored and controlled the soil and plant parameters in greenhouses with robots.
In the study by \cite{dt_eppinger2021}, Siemens briefly demonstrated two application examples: ketchup production optimization, and a spray-drying process for producing milk powder. %

The potential added value of Digital Twins for food science has been acknowledged by \cite{Verboven2020} and \cite{dt_henrichs2022}) in their reviews, which focused on Big Data, cloud computing, the Internet of Things, and sensorization. A picture of Digital Twins as an augmented reality monitoring system in real time was envisioned. 
While mechanistic models were advocated, only simple models were expected to be suitable for Digital Twins. 
\cite{Verboven2020} noted that nowadays, complex physics-based models would not be solvable in real time.
The review by \cite{dt_henrichs2022} focused on the value chain and production planning at the shop floor level.
An example of this \enquote{scheduling} perspective on Digital Twins can be found in the article by \cite{dt_Koulouris2021}.

Based on the criteria in \autoref{tab:DTcriteria}, we find manuscripts that partially fulfill the features of Digital Twins.
Several studies (\cite{dt_defraeye2019,dt_tagliavini2019,dt_shoji2022}) on Digital Twins understand the latter as a simulation model for post processing recorded temperature profiles of physical counterparts. While presenting well-validated models and numerical simulations, especially bi-directional links, interactive decision making, and mirroring over the product life cycle have not yet been addressed. The necessity of real time simulations has been partially envisioned (\cite{dt_prawiranto2021}). However, the presented simulation speed up compared to real time of factor three is far too low for online decision making. The former can be explained by the lack of reduced-order modeling (ROM) in all of the studies mentioned above.

%
Several studies on reduced-order models have been published in food science. 
In a recent review by \cite{rom_khan2022}, only two of the 31 presented studies were ROMs of transient nature. The first was a contribution by \cite{mpc_Huang1998}, using neural networks to train simple 
n-step ahead predictors. 
The second was a contribution by \cite{mpc_Li2016}, where a recurrent self-evolving fuzzy neural network learns from simulation with errors of a maximum of \SI{2}{K}. Unfortunately, no information on the computational time were given in these studies.
The residual studies in \cite{rom_khan2022} revealed a \enquote{steady-state} perspective on machine-learning-based models in food science: experimental or simulation end results are used to train feed-forward neural networks, the processing time is treated as an input, and process settings cannot be changed individually within the run, though only per case. 

Notable exceptions are the works by \cite{mpc_Alonso2013} and \cite{rom_Rivas2013}.
A transient Fourier’s law partial differential equation is Galerkin-projected to a small subspace. One drawback of the employed proper orthogonal decomposition lies in its equation-invasive behavior (\cite{rom_Trehan2016,rom_Hesthaven2018}). The approach consequently excludes the use of commercial software, where typically most of the modeling is performed (\cite{Verboven2020}), and where root-level user-defined solver access is seldom.
In a more recent study, the authors stated that their solution times need to be reduced significantly for true real-time optimal control~(\cite{mpc_Alonso2021}). Their underlying optimization routine was enforced to stop early after five minutes to proceed with the next model predictive control time-step.

Several studies that aim at the optimal control of thermal processing do not use ROM, and instead reduce the spatial resolution and topology of the model. \cite{mpc_Arias-Mendez2013} modeled a 2D quarter section of an axisymmetric potato chip with 200 elements, which requires \SI{40}{s} to simulate \SI{90}{s} of real time. \cite{mpc_Hadiyanto2008} utilized 10 1D elements resulting in 108 ordinary differential equations that were solved in \SI{4243}{s} for \SI{9000}{s} of real time. \\


The outline of this article is as follows.
The physics-based data-driven Digital Twin framework for autonomous thermal food processing is proposed in \autoref{sec:ourdtframework}.
The focus of this paper lies on one essential building block of the framework: non-intrusive reduced-order modeling. 
%
%
The employed methods for reduced-order modeling are the physics-based food processing model (\autoref{sec:foodmodel}), the neural network-based ROM approach (\autoref{sec:DynROM}), and signal synthesis and ROM error measures (\autoref{sec:signalsynth}).
The main focus of \autoref{sec:results} is to answer the question of how to build high-fidelity, faster-than-real-time ROMs of food processing. The presented generation of a ROM is parsimonious in training signals, as only one data set is employed.
\autoref{sec:introexample} demonstrates the approach to derive six global test error measures from ROM evaluations over time. \autoref{sec:trainandeval} proposes how to evaluate trained ROMs on fair test data sets. 
\autoref{sec:inputspace} considers the influence of input space coverage on ROM's test errors.
\autoref{sec:corrs} investigates if excitation signal properties can be correlated to the test error -- to put it differently, to estimate the ROM accuracy a priori. This enables the a priori selection of a suitable single training signal to train a ROM.
\autoref{sec:signaltype} and \autoref{sec:sinAPRBS} attempt to find the best excitation signal type among the common multi-step or multi-sine signals in the literature.
 \renewcommand{\sectionautorefname}{Sec.}\renewcommand{\subsectionautorefname}{Sec.}\renewcommand{\subsubsectionautorefname}{Sec.}%
%
This study is one of the first scientific publications which employed the Dynamic ROM method. Consequently, evaluations over time, ROM speed-ups, computational cost, and ROM accuracy are investigated in~\autoref{sec:romevals}. 
%
 %
Following a summary and discussion of the main results in \autoref{sec:conclusionandoutlook}, this article discusses what benefits for food science can be drawn, broadening the perspective of how the approach contributes to a spectrum of applications in food science and beyond.
The outlook presents potential future steps to complete the proposed Digital Twin framework for processing autonomy.
\section{Methods} \label{sec:methods}

\subsection{A physics-based data-driven Digital Twin framework enabling autonomy of thermal food processing} \label{sec:ourdtframework}

\begin{figure}[htbp]
 \centering
\includegraphics[width=0.7\columnwidth]{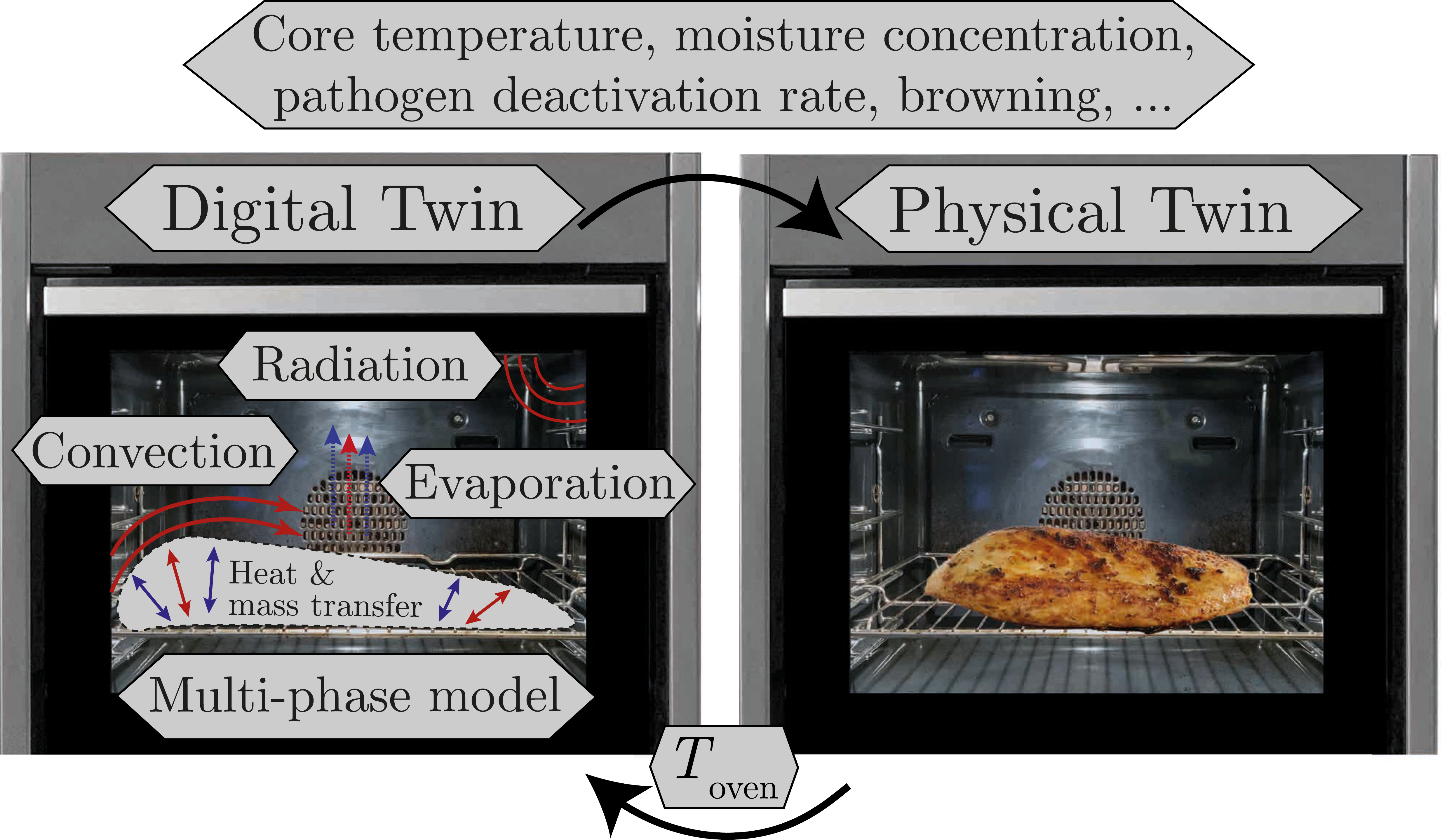}
 \caption{Digital Twins for autonomous cooking processes.}
 \label{fig:dt_oven}
\end{figure}

Presently, cooking devices cannot easily measure the food’s sensory properties, such as the chemical reactions (e.g., Maillard browning influencing color and flavor), texture (such as tenderness), or residual moisture content without user interaction. 
Here, the Digital Twin may provide the cooking product's unknown state information (\autoref{fig:dt_oven}).
Simple temperature probes inside the oven form the boundary and initial conditions of the simulation model. Faster-than-real-time simulations return multiple future scenarios to the device’s autonomous process control.
The device can plan its oven temperature trajectory to meet the user's requirements of specific brownness, moisture content, minimum core temperature, or pathogen deactivation at the end of the process -- at a specific point in time.
An example of a user input to an autonomous cooking device could be: \emph{Prepare safe medium-rare meat, tender, thorough browning and reach ready-to-eat temperatures at 08:15 p.m.} A change of the user's objectives is also possible during the process, just limited by the current progress of cooking. Reaction to disturbances, such as power loss or opening of device doors, is self-evident for model predictive control.

This study formulated nine requirements that the authors believe are necessary to accomplish autonomous processing at the device level (\autoref{tab:DThypothesis}).
\begin{table*}[tbhp]
\scriptsize
\centering
\addtolength{\tabcolsep}{-0pt}
 \caption{Hypotheses on requirements for a Digital Twin enabled process autonomy.} 
\begin{tabularx}{\textwidth}{l l X}
\toprule 
\textbf{Requirement}        & \textbf{Relates to}     & \textbf{Description}            \\ 
\midrule                                                                                                                                                                                                                                                                                                                                                                                                                                                                                                                                                                                                                                                                                               
R1: Faster than real-time                               & F2             & Digital Twin process autonomy requires faster-than-real-time simulations.                                                                                                                            \\
R2: Physics-based modeling                       & F3             & First principle models provide profound understanding and subsequently enable control over the process.                                                                            \\
R3: On-device operation   & F2, F3         & For execution at the device level, the computational load of the simulation has to be minimal.                                                                                                     \\
R4: Cybersecurity                                       & F5, F7, R3     &On-device operation increases cyber-security, especially for automation of system-critical processes.              \\ 
R5: Minimum physical sensors                             & F1, F5,        & Focus on easy-to-measurable quantities that are used as boundary conditions for the simulation.     \\
R6: Intellectual property protection                & F3, F4, F5, R3 & Modeling know-how is the core intellectual property of companies and should not be left unprotected.       \\
R7: Non-intrusiveness of ROMs                              & F3, F4, F6, R2 & There is seldomly code access to solvers in commercial simulation software.\\ 
R8: Data determines ROM quality &  F3, F7, R7     & Data-based ROM methods require sophisticated excitation signal design to accomplish ROM accuracy.              \\ 
R9: Open access                                        &   F6             & The distribution of Digital Twins should be open-source code-based to enable an application at scale.   \\
\bottomrule 
 \label{tab:DThypothesis}                                                                                                                                                                                                                                                                                                                                                                           
\end{tabularx}
\end{table*}
Model predictive control is only attainable with simulation times that are several magnitudes faster than real-time simulations (R1), as the models are called several hundreds of times during one control step (\cite{mpc_Arias-Mendez2013}). Instead of following the Big Data trend, this study aimed to provide fast and yet highly accurate (F3) virtual replicates of physics-based (R2) models, with parsimonious training data input (R8).

Non-intrusiveness (R7) of ROMs is essential for an industrial application at scale. Instead of using high-performance computing, the authors believe that Digital Twins must become very lean in data storage and computational power requirements. Digital Twins should be executable on the device-level (R3). It is not feasible or sustainable to execute the on-device simulations on high-performance clusters from an energetic viewpoint.
Exploiting existing standard sensors renders the product affordable, as complex sensors can be replaced by virtual probes (R5).

\begin{figure*}[hbtp]
 \centering
\includegraphics[width=0.82\textwidth]{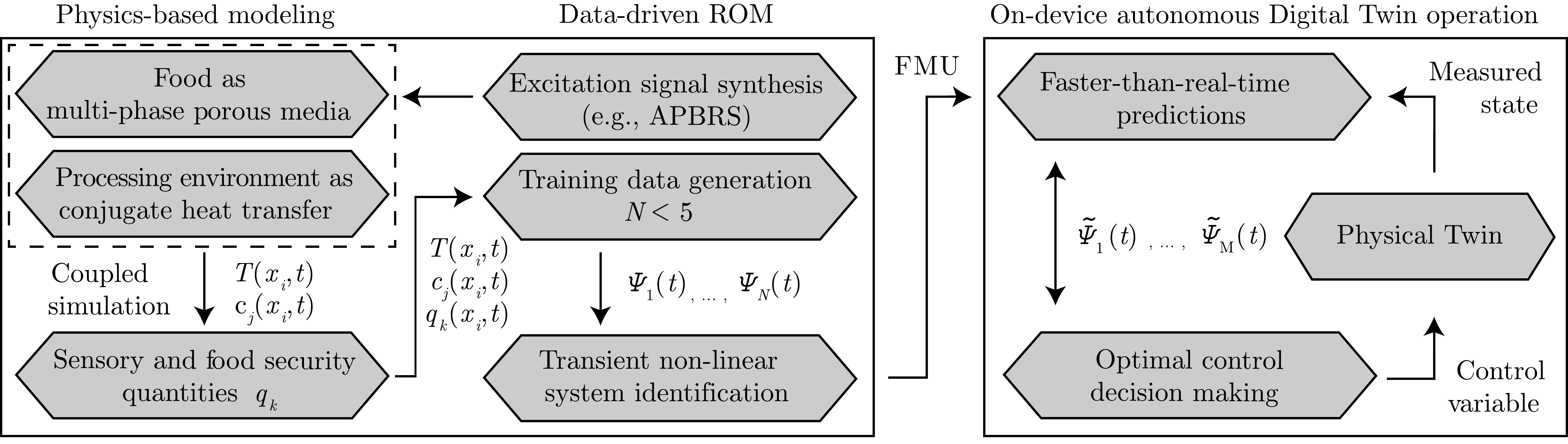}
 \caption{Concept of the proposed physics-based data-driven Digital Twin framework.}
 \label{fig:dt_famework22}
\end{figure*}
The state of the art in food science has not yet addressed many of the above-mentioned issues. It was the aim of our research initiative to close several such gaps, proposing a physics-based data-driven Digital Twin framework for autonomy. It fulfills the requirements of \autoref{tab:DThypothesis} with the following methods:
physically detailed full-order models of the process and product generate high-quality training data (see the left box of \autoref{fig:dt_famework22}). 
Conjugate product and process modeling entail thermal computational fluid dynamics (CFD) simulations coupled to porous media models. Temperatures $T$ and moisture concentrations $c$ are probed at various positions in the foodstuff. Temperature-dependent sensory attributes $q$ are solved subsequently.
All output data can be summarized within a result matrix $\Psi (t) = [T, c, q]$.
A non-intrusive, data-driven, non-linear, transient ROM method compresses the model to provide faster-than-real-time simulations.
The open-access (R9) model exchange file, termed \enquote{functional mockup unit} (FMU), contains the identified ROM in an encrypted container (R6). The file consisting of several megabytes is transferred to the device and executed live to produce estimates $\tilde{\Psi} (t)$ of the state with minimum system requirements (see right box of \autoref{fig:dt_famework22}).
Model predictive control algorithms can perform on-device decision-making.

\subsection{Food processing simulation model}\label{sec:foodmodel}
\begin{figure*}[htbp]
\centering
 \begin{subfigure}[]{0.59\textwidth} 
  \centering
\includegraphics[width=\textwidth]{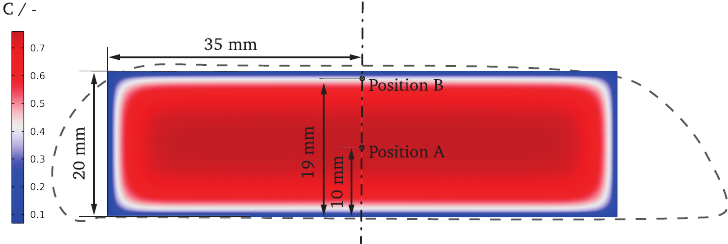} 
 \caption{Dimensions of the simulated chicken cuboid. Water mass fraction $C$ after \SI{1200}{s}.} \label{fig:RabelerCub-a}
 \end{subfigure}
  \begin{subfigure}[]{0.35\textwidth}
  \centering
  \setlength\fheight{0.18\textheight}
\setlength\fwidth{\textwidth}
\input{tiks/ROM-signal/RabelerCube-Set3-Valid}
 \caption{Validation of the implementation.} \label{fig:RabelerCub-b}
 \end{subfigure}
 \caption{Domain measures and validation of this simulation with experimental data of \cite{cm_rabeler_mod2018}.}
 \label{fig:RabelerCube-Set3-Valid}
\end{figure*}

This study considers the processing of chicken fillets in a convection oven. 
One usual modeling concept for the thermal processing of foodstuff is the porous media approach~(\cite{pm_datta_porous2007a}). Successful implementation and reduced-order modeling have been demonstrated lately (\cite{diss_eccomas21}). Regarding meats, the soft matter approach has been successfully applied~(\cite{cm_vandersman2007a}).
This study implemented a variant by~\cite{cm_rabeler_mod2018}, as the researchers provided matching sensory Arrhenius equations in subsequent studies.
It assumes a water-saturated foodstuff, the absence of a gas phase, and capillary pressure is considered zero~(\cite{pm_Datta2012}). Hence, only the conservation equation for mass (moisture concentration $c$) and energy (temperature $T$)
\begin{linenomath*}
\begin{align}
\label{eq:conservation-differential}
\frac{\partial c}{\partial t} + \frac{\partial}{\partial x_i} \left(c  u_i  \right) &= \frac{\partial}{\partial x_i} \left(D_\text{cb} \frac{\partial c}{\partial x_i} \right)\,, \\
\label{eq:energy}
\rho c_\text{p}  \frac{\partial T}{\partial t} + \rho_\text w c_\text{p,w} u_i \frac{\partial T}{\partial x_i}  &=  \frac{\partial}{\partial x_i} \left(\lambda_{ij} \, \frac{\partial T}{\partial x_j}\right)\,,
\end{align}
\end{linenomath*}
are solved, where $u_i$ denotes water velocity. $D_\text{cb}$ is the moisture diffusivity, and the orthotropic heat conductivity $\lambda_{ij}$ accounts for the fiber direction of the chicken fillet, 
$c_\text{p}$ is the effective heat capacity, $\rho$ is the effective density of chicken. The additional index w marks the corresponding parameters for water. 
The dimensions of the cubic chicken fillet, as depicted in \autoref{fig:RabelerCub-a}, follow the experimental setup by \cite{cm_rabeler_mod2018}. This facilitates the validation of model implementation.
Along the surface of the foodstuff, a heat transfer coefficient $h_\text{amb}$ is applied to model the surrounding convection and radiation effects related to the oven temperature $T_\text{oven}$. The evaporation rate of moisture $m_\text{evap}$ acts as a mass and latent heat ($H_\text{evap}$) loss in the boundary conditions for mass and energy:
\begin{linenomath*}
\begin{align}
  \vect{q_\text{mass}} \cdot \vect{N} &= -\frac{m_\text{evap}}{M_\text{w}} \,,  \\
\vect{q_\text{energy}} \cdot \vect{N}  &=  h_\text{amb}(T_\text{oven}-T) - m_\text{evap} H_\text{evap}\,,
\end{align}
\end{linenomath*}
where $\vect{q_\text{mass}}$ and $\vect{q_\text{energy}}$ are the surface mass and energy fluxes. $\vect{N}$ is the normal vector on the surface, and $M_\text{w}$ is the molecular weight of water.
The derivation of these boundary conditions, the swelling pressure model for moisture convection speeds $u_i$, and the utilized material properties can be found in~\ref{sec:chickenmodel}.
This study focuses on core temperatures $T_\text{A}$ and surface temperature $T_\text{B}$. 
Their positions are marked in \autoref{fig:RabelerCub-a}. Core temperatures typically serve as the primary parameter to control doneness. For example, \cite{mot_foodcode2017} requires core temperature holding times of at least one second above \SI{74}{\degreeCelsius}. Surface temperatures are relevant for determining temperature-dependent sensory parameters.

\paragraph{Discretization and solution procedure}The model was implemented in the \textit{Coefficient Form PDE} module of COMSOL Multiphysics 5.6. 
2528 hexahedral Lagrange elements, with a maximum height of \SI{2.75}{mm} were used to discretize the quarter section model of a cuboid ($70 \times 20 \times \SI{40}{mm}$), as depicted in \autoref{fig:RabelerCub-a}. Four inflation layers (growth-factor 1.3, \SI{0.2}{mm} first layer height) have been added to capture subsurface gradients. A relative discretization error  of \SI{0.17}{\%} compared to the grid-independent solution was estimated with generalized Richardson extrapolation by \cite{grid_roache}.
The resulting system of equations has \SI{51094}{} degrees of freedom. Time-stepping is performed with adaptive second-order backward differencing formulas, with a maximum time step of \SI{10}{s}. One simulation run takes \SI{9.5}{min} on four cores of a Core i9 processor.
This implementation's root mean square validation error was $\text{RMSE}=\SI{2.86}{K} $ for core temperatures $T_\text{A}$ and $\text{RMSE}=\SI{1.52}{K} $ for surface temperatures $T_\text{B}$, as depicted in \autoref{fig:RabelerCub-b}.

\subsection{Dynamic data-driven ROM}\label{sec:DynROM}
Recent advances in neural network-based ROMs promise to cover non-linear behavior of models with high precision.
The software package \textit{Dynamic ROM Builder}\footnote{\copyright~2020 ANSYS Inc.; \cite{rom_DynROMPatent}.} (termed DynROM hereafter) can provide a non-intrusive, non-linear, transient ROM of the full order system simulations of core ($T_\text{A}$) and surface temperatures ($T_\text{B}$) of the chicken fillet. This study considers these two points to be representative of the range of possible temperature trajectories. The solution matrix of dimension $n \times N$ reads $\vect{Y} =  \left[ T_{\text{A},1} , ... , T_{\text{A},N}; T_{\text{B},1} , ... , T_{\text{B},N} \right]$ with $n=2$ system states over $N=280$ time steps with $\Delta t = \SI{5}{s}$.
The ROM is based on an ordinary differential equation (ODE) system equation
\begin{linenomath*}
\begin{align}\label{eq:DynROMODE}
 \begin{pmatrix} \vect{X}' \\ \vect{I}' \end{pmatrix} &= f \left( \begin{pmatrix} \vect{X}\\ \vect{I}\end{pmatrix}, \vect{U},\vect{X}_0 \right)\,,\\
    \vect{X}(t=0) &= \vect{X}_0 \,,
\end{align}
\end{linenomath*}
where 
$\vect{U}$ is the vector of system excitations (oven temperatures) $\vect{U} = \left[ T_{\text{oven,}1} , ... , T_{\text{oven,}N} \right]$
and 
$\vect{X}$ is the simulation state vector equivalent to $\vect{Y}$. 
$f()$ is modeled with a three-layer feed-forward neural network, sigmoid activation functions, and is trained with gradient descent optimization~(\cite{rom_calka2021}). 
A novel concept is the addition of $i$ free state variables $\vect{I}$ (dimension $i \times N$) in the state vector. 
This ensures learning of the system’s non-linearities~(\cite{rom_twinbuilder_2020}). The vector can be interpreted as memory cells that are added as long as it increases the model quality. 
The ODE system is integrated with fourth-order Runge Kutta schemes.
The loss function for the neural network training error is the mean squared error 
\begin{linenomath*}
\begin{align}
    \text{MSE} =  \frac{1}{n} \sum_{l=1}^n \left(\frac{1}{N} \sum_{j=1}^{N} (X_{l,j} - Y_{l,j})^2 \right)
\end{align}
\end{linenomath*}
which is averaged for all learning scenarios.
%

\subsection{Signal synthesis for a parsimonious design of experiment} \label{sec:signalsynth}
When employing a data-based, non-intrusive ROM approach, the focus shifts to a \enquote{best-possible} data synthesis, as this is the only information vehicle for the subsequent system identification. The ROM test error is strongly connected to the data presented during training~(\cite{sig_Heinz2017,sig_Kenett2022}).
Only a few studies on correlations between the excitation signal and ROM learning success can be found, especially for machine-learning-based ODE ROMs.
However, \cite{sig_Gringard2016} and \cite{sig_Heinz2017,sig_Heinz2018} provide suggestions for the design of experiment for non-linear auto-regressive exogenous models and local model networks, which will be used as guidance hereafter. 

This section presents the signal synthesis for typical excitation signals within the literature.
Typically, two classes of excitation signals -- sinusoids and multi-steps -- are employed for system identification. 
\cite{Nelles2020} promotes the usage of sinAPRBS, which can be interpreted as a combination of both (compare \autoref{fig:Time-sinAPRBS312a} for an example). 

\paragraph{Multi-level signals}
While step responses and pseudo-random binary sequences are commonly used for linear system identification, they are unsuitable for non-linear system identification~(\cite{Nelles2020}). Amplitude modulated pseudo-random binary sequences (APRBS) are expected to cover the input space adequately (\cite{sig_Heinz2017}). 
Jumps in the signal excite a broad range of system frequencies, and the piecewise constant sections cover low-frequency components.
APRBS signals are synthesized pseudo-randomly (compare \autoref{ALG_synapbrs}).
A minimum hold time $t_\text{hold} = \SI{300}{s}$ is set to respect the most considerable time constant of the model. 
{
\center{
\begin{minipage}{0.95\linewidth}
\begin{algorithm}[H]
\caption{Synthesis of APRBS signals for $T_\text{oven}$.}\label{ALG_synapbrs}
\footnotesize
\begin{algorithmic}[1]
\State Discretize the range of temperature between $T_\text{min} = \SI{293.15}{K}$ and $T_\text{max} = \SI{473.15}{K}$ into four equidistant bands
\State Compute $T_{\text{oven,}i}$ randomly in every $i$\textsuperscript{th} temperature band with $i=1,\ldots,4$ and permute $T_{\text{oven,}i}$

\State Compute temperature difference $T_\text{diff}$ between adjacent $T_{\text{oven,}i}$

\If{any $T_\text{diff} < T_\text{margin} = \SI{10}{K}$}
\State Go to step $\textbf{2}$ 
\EndIf

\State Compute time-steps $t_{i}$ randomly between $t_{start}$ and $t_{end}$ with $i=1,\ldots,4$
\State Compute time difference $t_\text{diff}$ between two adjacent time steps 
\If{any $t_\text{diff} < t_\text{hold}$}
\State Go to step $\textbf{7}$
\EndIf 
\State Sample $t_{i}, T_{i}$ over time with $\Delta t = \SI{5}{s}$ and set $T_{\text{oven,}0} = \SI{279.15}{K}$
\end{algorithmic}
\end{algorithm}
\end{minipage}
}
\par
}
\paragraph{Sinusoidal APRBS signals}
Following a proposition by \cite{sig_Heinz2016}, sinusoidal pieces $A \sin (2 \pi f_{j} t) +  c$  with frequencies $f_j \in [0.001,0.01]\,\SI{}{Hz}$ are inserted in APRBS signals to synthesize sinAPRBS. The amplitudes $A$ and the offset $c$ are adjusted to fit a quarter sine wave into each APRBS step.

\paragraph{Multi-sines signals}
Multi-sines are a superposition of several single sines:
\begin{linenomath*}
\begin{align}
    U(t) = \sum_{j=1}^n A \cos ( 2 \pi l_j f_0 t + \varphi_j) + c \quad \text{with} \quad l_j \in \mathbb{N}\,.
\end{align}
\end{linenomath*}
Parameters $f_0 \in [0.001,0.01]\,\SI{}{Hz}$, $\varphi_j \in [0,10]$ and $A \in [1,30]\,\SI{}{K}$ were adjusted to match the frequency and amplitude ranges of interest. 
Occasionally, the excitation signal's crest factor $ \operatorname{Cr}(\vect{U}) = \operatorname{max} |U_k| / U_\text{RMS}$, a measure of the energy level of the signal, is optimized with the usage of Schroeder phases (\cite{sig_Gringard2016}): 
\begin{linenomath*}
\begin{align}
\varphi_{j} = -\frac{j \,(j-1) \,\pi}{m} \quad \forall j \in [1,..., m]\,,
\end{align}
\end{linenomath*}
where $k$ marks a certain time step and $m$ is the number of excited sines.

During the benchmark case (\autoref{fig:RabelerCube-Set3-Valid}) at full load, a core temperature of \SI{74}{\degreeCelsius} was exceeded during a run of \SI{1200}{s}. Considering that multi-step and multi-sine signals do not apply full load at all times, the duration of the signals in this study is prolonged to \SI{1400}{s}.

\paragraph{KPIs of excitation signals}

This study's objective was to find properties of a training signal (termed KPIs hereafter) that could correlate with the test error of the ROM that was trained with the same signal. The oven temperature's crest factor $\operatorname{Cr}(\vect{U})$ 
was obtained with the formula introduced previously. 
Intuitive signal properties were standard deviation and mean of $T_\text{oven},\,T_\text{A}$ and $T_\text{B}$. 
Moreover, this study calculated various mean and absolute measures of the APRBS steps of \autoref{ALG_synapbrs}: $\overbar{T_i}$, $\overbar{T_{\text{diff},i}}$, $\overbar{T_{\text{diff},j}}$ , and $\overbar{| T_{\text{diff},i} |}$, with $i=1,\ldots,4$ and $j=2,\ldots,4$, hence excluding the first step from the calculation for $\overbar{T_{\text{diff},j}}$. $\overbar{T_{\text{diff},j}}$ is the signed mean of the delta jumps of an APRBS signal. The delta jumps $T_{\text{diff},j}$ are introduced in \autoref{ALG_synapbrs}.


\paragraph{Error measures} 

An accuracy measure to quantify the ROM test errors should be interpretable, equally fair concerning over- and under-predictions, scale-independent, insensitive to outliers and stable for zero crossings of the observation variable~(\cite{Chen2017}).
The root mean square error
\begin{linenomath*}
\begin{align}
    \text{RMSE} = \sqrt{\frac{1}{N} \sum_{k=1}^N (X_k - Y_k)^2}
\end{align}
\end{linenomath*}
is scale-dependent, facilitating interpretability. 
A percentage-based variant is the mean absolute percentage error 
\begin{linenomath*}
\begin{align} 
   \text{MAPE} = \frac{1}{N} \sum_{k=1}^N \frac{| X_k - Y_k |}{|Y_k|}\,.
\end{align}
\end{linenomath*}
Both are popular error measures as they are fair concerning over- and under-predictions. However, they demonstrate dominant sensitivity to single outliers. Due to the deterministic behavior of the model, single outliers were not expected in this study.
Moreover, MAPE is sensitive to zero crossings of the observation $y_k$ and
scale sensitive within one time series~(\cite{Chen2017}). 
In this study, all observations remained within an identical order of magnitude and had no zero crossings.
A local accuracy measure is the maximum absolute error
\begin{linenomath*}
\begin{align}
\text{MAX} = \mymax_{ k } | X_k - Y_k |\,.
\end{align}
\end{linenomath*}
Statistical measures of the time-dependent error were MEDIAN and IQR (interquartile range) as they are more robust to outliers than MEAN and STD (standard deviation). 
The models' goodness-of-fit was calculated with $\text{R}^2$ (coefficient of determination).

\section{Results and discussion} \label{sec:results}

\subsection{Parsimonious design of experiment for ROMs with one training signal -- Introductory example} \label{sec:introexample}

\begin{figure}[htb]
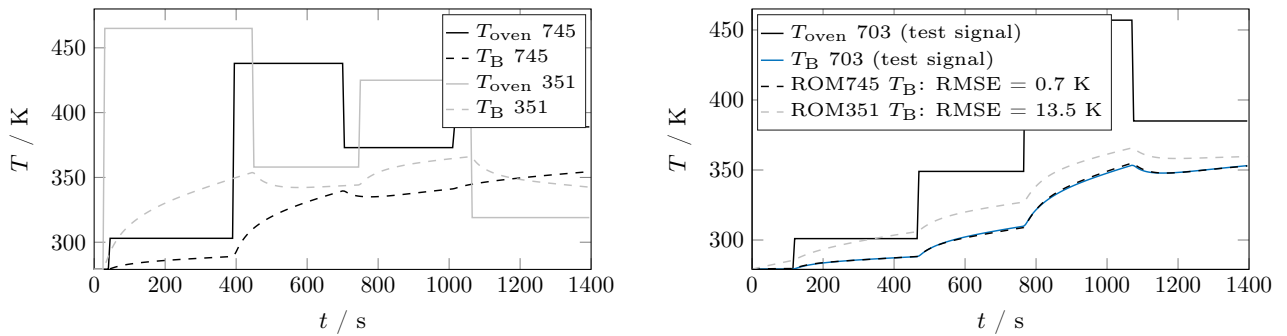

\centering
 \begin{subfigure}[]{0.49\columnwidth}
\centering
\setlength\fheight{0.22\textheight}
\setlength\fwidth{0.95\textwidth}
 \input{tiks/ROM-signal/Time-sinAPRBS745-Intro.tex}
 \caption{APRBS training signals 351 and 745.} \label{fig:Time-Introa}
 \end{subfigure}
 \begin{subfigure}[]{0.49\columnwidth}
  \centering
  \setlength\fheight{0.22\textheight}
\setlength\fwidth{0.95\textwidth}
   \input{tiks/ROM-signal/TimeEval-703AP-Intro1}
       \caption{Testing of ROM351 and ROM745 on test signal 703.}\label{fig:Time-Introb}
 \end{subfigure}
  \caption{Introductory example illustrating the sensitivity of signal selection for ROM training on the resulting test error.}
 \label{fig:Time-Intro}
\end{figure}

\begin{figure}[htbp]
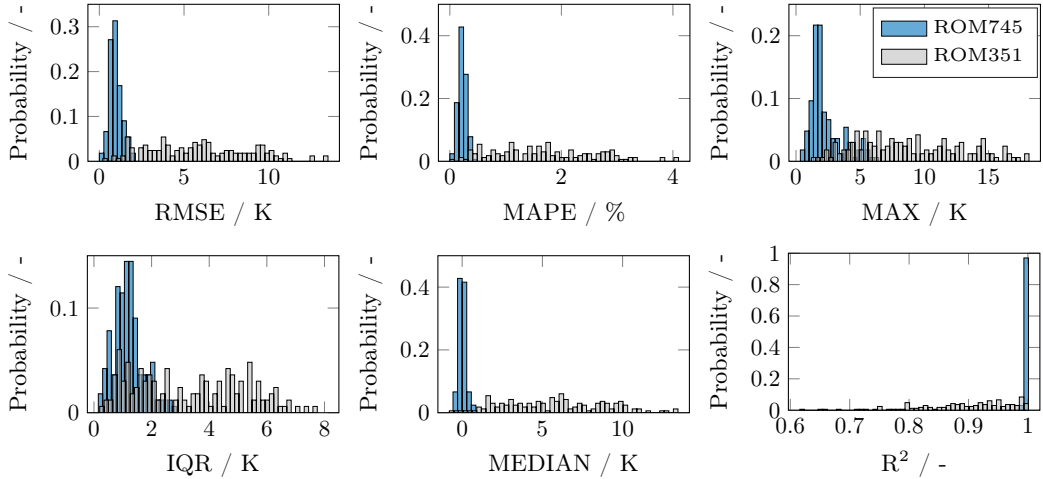

 \centering
\setlength\fheight{0.16\textheight}
\setlength\fwidth{0.28\columnwidth}
\input{tiks/ROM-signal/boxrms-745-rms}
\input{tiks/ROM-signal/boxrms-745-mape}
\input{tiks/ROM-signal/boxrms-745-max}\\
\input{tiks/ROM-signal/boxrms-745-iqr}
\input{tiks/ROM-signal/boxrms-745-median}
\input{tiks/ROM-signal/boxrms-745-r2}
    \caption{Distribution of error measures for 1-signal-ROM351 and for 1-signal-ROM745.}
    \label{fig:boxrms-745}
\end{figure}

In this first study part, one \enquote{best-possible} training signal is found. \autoref{fig:Time-Intro} illustrates the sensitivity of training signal selection with a short example. 
In \autoref{fig:Time-Introa}, two training signals with identifiers 351 and 745, are shown. Both signals are used individually to train a ROM. The testing is performed on signal 703 (compare \autoref{fig:Time-Introb}). Signal 745 produces good results with a RMSE (over time) of \SI{0.7}{K}.
Conversely, signal 351 produces a RMSE over time that is one order of magnitude larger.
Subsequently, ROM351 and ROM745 are evaluated on all available test signals to render testing more representative. Instead of demonstrating all solutions over time, for each evaluation, the test error measures RMSE, MAPE, MAX, MEDIAN, IQR and $\text{R}^2$ are calculated from $T_\text{B,Test} - T_\text{B,ROM}$ over all time steps. 
\autoref{fig:boxrms-745} shows the histogram of those error measures, calculated for all test data.
\begin{table}[htbp]
\centering
\scriptsize
\addtolength{\tabcolsep}{-2.1pt}
\caption{Average error measures of \autoref{fig:boxrms-745}.} 
\begin{tabular}{lrrr} \toprule 
& ROM351 & ROM745 \\ 
\midrule
RMSE / K   & 5.90 & 0.97 \\
MAPE / \%   & 1.68 & 0.22 \\
MAX / K    & 8.84 & 2.33 \\
MEDIAN / K & 5.67 & 0.01 \\
IQR / K     & 3.30 & 1.20 \\
$\text{R}^2$ / -      & 0.90 & 1.00 \\
\bottomrule 
 \label{tab:Eval on APRBSAllA4-Table-Foc-IntroComp 351-745} 
 \end{tabular}
 \end{table}

It appears that test error measures are normally distributed for a sufficiently large test data set.
It seems appropriate to average the error measures to obtain six global error measures per 1-signal-ROM.  \autoref{tab:Eval on APRBSAllA4-Table-Foc-IntroComp 351-745} displays the averaged error measures for ROM351 and ROM745.
Obviously, several signals convey more information on the physical system than others during reduced-order modeling. This study aims to find a property (KPI) of the signals that can allow us to select the \enquote{best-possible} training signal out of a large set of excitation signals. 

\subsection{ROM training and evaluation setup}\label{sec:trainandeval}

A total of 100 single sinAPRBS, APRBS, and multi-sine (with and without Schroeder phases) signals were used individually to train a ROM, termed 1-signal ROM hereafter. For better identification, the signals have a unique alphanumeric identifier ranging from 272 to 1087.
A fixed complexity of $i=2$ of the DynROM method ensures equal training conditions.
There is hardly any correlation ($R=0.096$) between training error ($\text{MEDIAN}(\text{RMSE}_\text{train}) = \SI{0.22}{K}$, $ \text{STD}(\text{RMSE}_\text{train}) = \SI{0.14}{K}$) and test errors. Consequently, the training error has no significant effect on the study outcomes with respect to test RMSEs of the ROMs.

The ROMs' training and testing have to follow an objective study design to render the study outcomes statistically significant. One finding is that testing has to be performed on fair evaluation sets introduced hereafter.%
\begin{figure}[thbt]
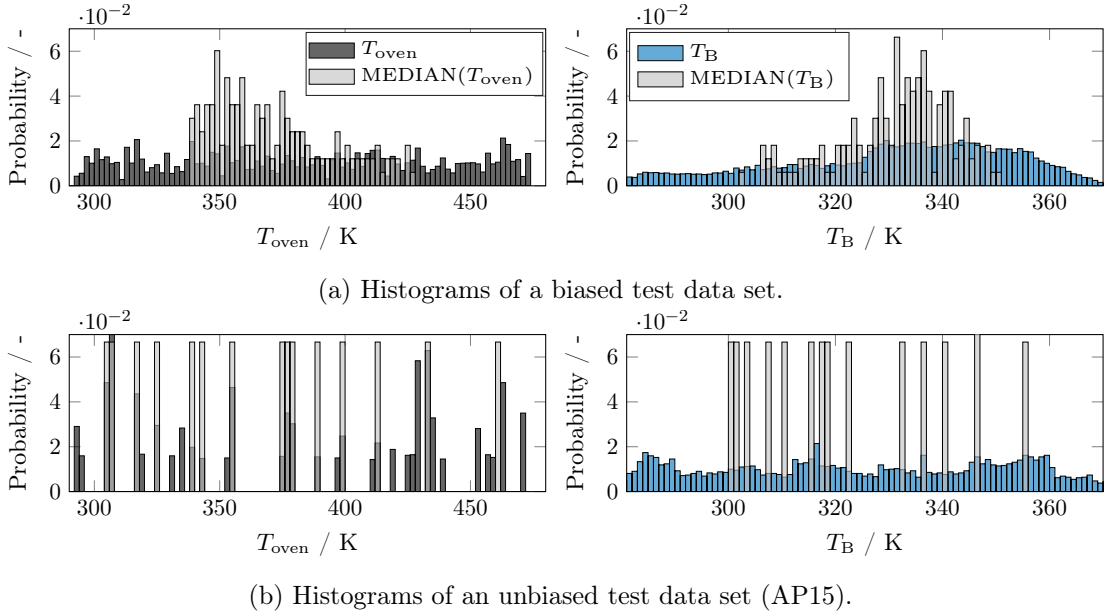

 \centering
  \begin{subfigure}[]{\columnwidth}
  \centering
\setlength\fheight{0.16\textheight}
\setlength\fwidth{0.45\textwidth}
\input{tiks/ROM-signal/chosenAllA4-histTin2} 
\input{tiks/ROM-signal/chosenAllA4-histTb2}
       \caption{Histograms of a biased test data set.}\label{fig:chosenall-Tb-hist-a}
 \end{subfigure}
 \begin{subfigure}[]{\columnwidth}
  \centering
\setlength\fheight{0.16\textheight}
\setlength\fwidth{0.45\textwidth}
\input{tiks/ROM-signal/chosenTb15-histTin2}
\input{tiks/ROM-signal/chosenTb15-histTb2}
       \caption{Histograms of an unbiased test data set (AP15).}\label{fig:chosenall-Tb-hist-b}
 \end{subfigure}
 \caption{Test sets with equally distributed $T_\text{oven}$ excitation signals (top left) show pronounced biased surface temperatures for testing (top right). Fair evaluation sets, here AP15, show a more uniform distribution of $T_\text{B}$ (bottom row).}
 \label{fig:chosenall-Tb-hist}
\end{figure}
It is common practice to test ROMs on input data sets that have not been used for training. The ultimate goal is to predict how the estimated ROM can generalize on a yet unknown data set.
Cross-validation can be considered a method to separate data into train and test sets with multiple iterations and combinations. 
One possible approach is the k-fold cross-validation, where k is the number of portions into which the original data set is divided.
The presented procedure can be considered the limit case for a k-fold cross-validation with $k=N_\text{total}$: all signals were used individually to train a ROM and were evaluated upon complementary signals.

One central requirement for ROMs is validity in a defined variable range. Hence, test data should objectively represent this range.  
However, pseudo-randomly distributed APRBS excitation signals $T_\text{oven}$ show a normal-like distribution in their medians. Due to the physical nature of the heating process, the distribution can also be observed for $T_\text{B}$ (compare ~\autoref{fig:chosenall-Tb-hist-a}). 
The first plot shows a histogram of all amplitudes of $T_\text{oven}$ and the median oven temperature MEDIAN($T_\text{oven}$) per signal. The second plot depicts a histogram of all values of $T_\text{B}$ and the MEDIAN($T_\text{B}$) per signal.
One can visually consider the pseudo-random oven temperature order as a random experiment with two or more dice. The sum has a normal distribution.

It was discovered that leaving only the rest of the data set as test data would not produce neutral feedback on the ideal characteristics of an excitation signal. 
The test data would advantage ROMs that were trained with signals with a MEDIAN($T_\text{B}$) that is close to the mean median of all $T_\text{B}$.
The bias holds for random picks out of the complementary data set as well.
The concept of befitted cross-validation motivates to include expert input concerning the data generation structure during the selection process for test data sets~(\cite{sig_Kenett2022}). In that spirit, one may postulate that fair test data sets should be uniformly distributed concerning the medians of the temperature range of $T_\text{oven}$.
With a null hypothesis, \enquote{our observations of $T_\text{B}$ come from a uniform distribution}, this study selected test set candidates methodically, with a $\chi^2$ test ($p=0.05$, $n_\text{bins}=6$) for each class of signal types. The test sets with 15 signals per set are termed AP15 (APRBS), sinAP15 (sinAPRBS), and MS15 (multi-sines) hereafter.
Fair evaluation sets, such as AP15, reveal an equal distribution of $T_\text{B}$ and evenly distributed median temperatures per signal case~(\autoref{fig:chosenall-Tb-hist-b}).

Typically, the steps of $T_\text{oven}$ in APRBS signals cannot be realized by convection ovens instantaneously due to \enquote{lagging} thermodynamical effects.
Hence, sinAPRBS signals are more suitable to emulate this behavior during testing. 
Moreover, a mixed test set was created by adding AP15, sinAP15, and MS15, to emulate a mixed operational usage of the model. The test sets are presented in the supplementary material of this article. 
Each ROM is tested on AP15, sinAP15 and MS15. The average error measures are calculated individually for each test data set.
Note that none of the hereafter discussed training signal candidates are included in the test sets. This would, of course, bias the results.

\subsection{The influence of input space coverage on test error} \label{sec:inputspace}

\begin{figure*}[htb]
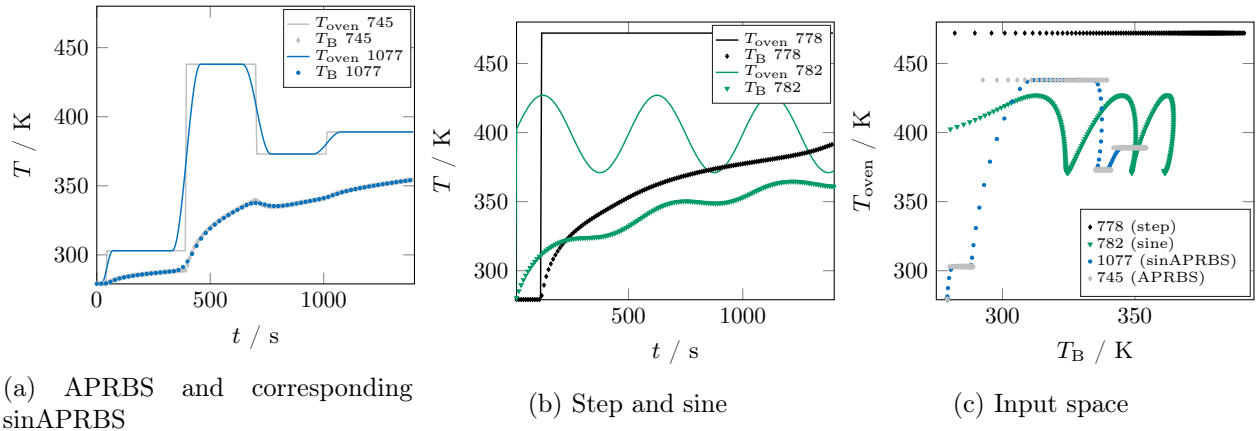

\centering
\setlength\fheight{0.23\textheight}
\setlength\fwidth{0.33\textwidth}
 \begin{subfigure}[]{0.31\textwidth}
 \input{tiks/ROM-signal/Time-sinAPRBS745.tex} 
 \caption{APRBS and corresponding sinAPRBS} \label{fig:Time-sinAPRBS312a}
 \end{subfigure}
  \begin{subfigure}[]{0.31\textwidth}
   \input{tiks/ROM-signal/Time-const778-sine782.tex} 
       \caption{Step and sine}\label{fig:stepsine}
 \end{subfigure}
 \begin{subfigure}[]{0.3\textwidth}
   \input{tiks/ROM-signal/778-input-combine.tex} 
       \caption{Input space} \label{fig:Time-sinAPRBS312c}
 \end{subfigure}
  \caption{Training cases with step, sine, APRBS and sinAPRBS signals and their corresponding input space.}
 \label{fig:Time-sinAPRBS312}
\end{figure*}

Often, literature suggests a uniform coverage of the models' input space (\cite{Nelles2020}) or output space (\cite{sig_Talis2021}) when synthesizing excitation signals. The input space is spanned by the excitation (here: $T_\text{oven}$) and system states (here: $T_\text{A}$ or $T_\text{B}$). Excitation signal and system states serve as inputs to the neural network of DynROM (compare \autoref{eq:DynROMODE}). One expects to cover the maximum of the system’s operating conditions while equally covering the input space. As the system states depend on past states, an input space covering signal design is a challenging task~(\cite{sig_Heinz2017}).
\autoref{fig:Time-sinAPRBS312} visualizes the concept of input spaces. The simulation input and output are recorded over time for an APRBS (745) and a sinAPRBS (1077) excitation (\autoref{fig:Time-sinAPRBS312a}) together with a step (778) and single-sine (782) excitation (\autoref{fig:stepsine}).
The excitation signal $T_\text{oven}$ and system output $T_\text{B}$ can be related to the points in the input space plot of \autoref{fig:Time-sinAPRBS312c}. An optimal input space coverage would implicate a uniform, two-dimensional distribution of points in this figure.
The implemented input space coverage measure ($\mathrm{Cv}$) in this article is inspired by \cite{sig_Heinz2018}, which is based on the maximin criterion (\cite{Johnson1990}):
\begin{linenomath*}
\begin{align}
   \mathrm{Cv} &= \frac{1}{N} \sum_{k=1}^N \mathrm{dist}(\tilde{\vect{Z}},Z_k)\,,\\
    \mathrm{dist}(\tilde{\vect{Z}},Z_k) &= \mymin_{m} | Z_m - Z_k |\,,
\end{align}
\end{linenomath*}
where $\tilde{\vect{Z}}$ is the input space matrix $\vect{Z} = [\vect{U}' ,\,\vect{Y}']$ without the currently considered row (time-step) $Z_k = [ T_{\text{oven},k}\, T_{\text{B},k}] $. 
Note that only signals with equal number of time-steps $\mathrm{card}(\vect{Z})=N$ are comparable with this implementation.
The $\mathrm{dist}()$ function is realized via nearest neighbors with Euclidian distances. A high value of $\mathrm{Cv}$ implies that areas without data points are penalized best.

\autoref{tab:Eval on sinAPRBSoldChi15-Table-Foc-IntroSet} summarizes the input space coverage $\operatorname{Cv}$ of the example signals in \autoref{fig:Time-sinAPRBS312}.
 \begin{table}[htbp]
 \scriptsize
\centering
\caption{Introductory example: comparison of input space coverage $\operatorname{Cv}$ and 1-signal-ROM test errors for step, sine, APRBS and sinAPRBS signal types.} 
 \begin{tabular}{lrrrrr} \toprule 
ROM-ID & ROM778 & ROM782 & ROM745 & ROM1077 \\ 
Signal type & step & sine & APRBS & sinAPRBS \\
\midrule 
Cv / - & 0.39 & 1.07 & 0.27  & 0.97\\
\midrule 
RMSE / K  & 24.09 & 2.95 & 1.31 & 3.69 \\ 
MAPE / \% & 5.68 & 0.75 & 0.28  & 0.84  \\ 
MAX / K& 44.16 & 6.00 & 3.02 & 7.28  \\ 
MEDIAN / K  & -19.81 & 2.24 & 0.25  & 0.78  \\ 
IQR / K  & 24.33 & 2.84 & 1.64 & 4.97  \\ 
$\text{R}^2$ / -  & -0.32 & 0.98 & 1.00  & 0.95  \\ 
\bottomrule 
 \label{tab:Eval on sinAPRBSoldChi15-Table-Foc-IntroSet} \end{tabular}
 \end{table}
Moreover, it shows the test errors on sinAP15 when each signal is used to train a 1-signal-ROM.
Although the APRBS signal possesses the lowest input space coverage, it has the lowest test error.
It becomes evident how careful signals should be picked for non-linear ROM training. Although all signals show a similar trajectory of $T_\text{B}$, a significantly different test success can be observed.
Single-step or sinusoidal signals should certainly not be employed for ROM training. 
APRBS seem to convey more information to the ROM than their sinAPRBS counterparts. sinAPRBS signals increase the ROM test error compared to their APRBS counterpart signal.
Regarding the presented thermal processing of chicken meat, the directive of maximizing the input space coverage might -- surprisingly -- not hold. Although the single sine provides the highest input space coverage Cv in this introductory example, it by far does not provide the lowest test error (see \autoref{tab:Eval on sinAPRBSoldChi15-Table-Foc-IntroSet}).
In the following section, the presented findings are investigated in a more statistically representative setting.

\subsection{Correlations between signal KPIs and test errors}\label{sec:corrs}
This section investigates correlations between signal properties (KPIs) and ROM test errors. Such information would allow to synthesize and select the \enquote{best-possible} excitation signals that ensure low ROM test errors -- just based on measures that can be calculated from the training data.
One can differentiate a priori (only evaluating the excitation signal) and a posteriori (evaluating excitation signal or system output or both) KPIs .

\begin{figure*}[htbp]
\raggedleft
\setlength\fheight{0.1\textheight}
\setlength\fwidth{0.25\textwidth}
 \begin{tabular}{p{0.31\textwidth}p{0.31\textwidth}l}
   \input{tiks/ROM-signal-evals/APRBSEqual15-RMSE-stdTb-Showsr5nl473f0005-a4-m1-m2-m3-m4-m5-m6t-m7-m8-m9-m10schroeder_mod}
 &
\input{tiks/ROM-signal-evals/sinAPRBSoldChi15-RMSE-stdTb-Showsr5nl473f0005-a4-m1-m2-m3-m4-m5-m6t-m7-m8-m9-m10schroeder_mod}
 &
\input{tiks/ROM-signal-evals/MultiSineEqual15-RMSE-stdTb-Showsr5nl473f0005-a4-m1-m2-m3-m4-m5-m6t-m7-m8-m9-m10schroeder_mod} 
\\
\input{tiks/ROM-signal-evals/APRBSEqual15-RMSE-mean-d-Ai-2-Show-a4_mod}
 &
\input{tiks/ROM-signal-evals/APRBSEqual15-RMSE-Lmean-Show-a4_mod}
  &
\input{tiks/ROM-signal-evals/sinAPRBSoldChi15-RMSE-Lmean-Showsr5nl473f0005-a4-m1-m2-m3-m4-m5-m6t-m7-m8-m9-m10schroeder_mod}
\end{tabular}
 \caption{Correlations of signal KPIs  and error measures for APRBS ( \protect\includegraphics[scale=0.4]{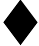} ), Best 5 APRBS ( \protect\includegraphics[scale=0.4]{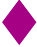} ), sinAPRBS ( \protect\includegraphics[scale=0.37]{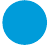} ), Multi-Sine ( \protect\includegraphics[scale=0.4]{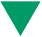} ), Schroeder Multi-Sine ( \protect\includegraphics[scale=0.37]{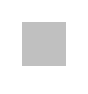} ). Regression curve and $p=0.95$ bounds for APRBS signals are added when appropriate. } 
 \label{fig:KPIandErrorCorrelations}
\end{figure*}

Concerning APRBS signals, the most significant correlation could be found for STD($T_\text{B}$), which is an a posteriori KPI. The top row of~\autoref{fig:KPIandErrorCorrelations} depicts the correlation between STD($T_\text{B}$) and test RMSE on the test sets AP15, sinAP15 and MS15. A regression curve fit (black line) and $p=0.95$ bounds (dashed lines) were added for the APRBS signals (black diamonds). Each marker symbolizes the evaluation of a 1-signal-ROM on the respective test set. The RMSE is the mean error for the test set, as explained in \autoref{sec:introexample}. Note that the strong STD($T_\text{B}$)-to-test-RMSE-correlation does not hold for multi-sine testing (top right plot of~\autoref{fig:KPIandErrorCorrelations}). 

A summary of all correlations can be found in \autoref{tab:Corr-Show-a4-Eval-APRBSEqual15}. It denotes Pearson's correlation coefficient between test error measures (rows) and a priori and a posteriori excitation signal KPIs (columns) for APRBS signals. Testing was performed on AP15. %
Correlation tables for the residual test sets sinAP15 and MS15 are given in the supplementary material of this article. 
The presented correlation trends hold not only for the data sets AP15, sinAP15 and MS15. The trends also hold for testing on all complementary APRBS signals and for other large sinAPRBS and APRBS data sets not presented in this article. %

The correlation between input space coverage Cv and error measures is low (compare respective column in \autoref{tab:Corr-Show-a4-Eval-APRBSEqual15}). It even reveals the opposite trend direction to what is expected from advice in the literature. $R=0.33$ decodes to: high input coverage results in high test errors. This is also visualized in the center of the bottom row in \autoref{fig:KPIandErrorCorrelations}. Moreover, although multi-sine signals have high input space coverages, they do not significantly correlate with the error measures (see the top row and bottom right plot of \autoref{fig:KPIandErrorCorrelations}).

\renewrobustcmd{\bfseries}{\fontseries{b}\selectfont}
\newrobustcmd{\B}{\bfseries}

\begin{table*}[htbp]
\scriptsize
\addtolength{\tabcolsep}{-3.1pt}
\sisetup{detect-weight,     
         mode=text,         
         table-format=-0.4, 
         add-integer-zero=false,
         table-space-text-post={*} 
         }
\centering
 \caption{Pearson's correlation matrix between error measures (rows) and a priori and a posteriori excitation signal KPIs (columns) for APRBS 1-signal-ROMs (test set: AP15). STD($T_\text{B}$) shows the best a posteriori correlation while $\overbar{T_{\text{diff},j}}$ is a potential a priori KPI.  } 
 \vspace{1em}
 \begin{tabular}{lrrrrrrrrrrrrrr} \toprule 
& $\overbar{T_i}$ & $\overbar{T_{\text{diff},i}}$ & $\overbar{T_{\text{diff},j}}$ & $\overbar{| T_{\text{diff},i} |}$ & $\overbar{T}_\text{oven}$ & STD($T_\text{oven}$) & Cr($T_\text{oven}$) & Cv($T_\text{oven}$) & Cv($T_\text{B}$) & Cv($T_\text{oven},T_\text{B}$)  & STD($T_\text{A}$)  & STD($T_\text{B}$) \\ 
\midrule 
RMSE    & 0.18            & -0.39                 & \B{-0.68}                & 0.40                    & 0.33                    & -0.12              & -0.13             & 0.25              & -0.29            & 0.33                         & -0.05             & \B-0.76           \\
MAPE   & 0.20            & -0.38                 & \B-0.69                 & 0.43                    & 0.35                    & -0.11              & -0.14             & 0.24              & -0.28            & 0.37                         & -0.06             & \B-0.78             \\
MAX   & 0.13            & -0.47                 & \B-0.69                 & 0.33                    & 0.25                    & -0.13              & -0.10             & 0.26              & -0.36            & 0.22                         & 0.03              & \B-0.70             \\
MEDIAN  & 0.19            & -0.37                 & \B-0.72                 & 0.50                    & 0.35                    & -0.09              & -0.13             & 0.19              & -0.32            & 0.43                         & -0.11             &\B -0.83             \\
IQR   & 0.31            & -0.07                 &\B -0.32                 & 0.23                    & 0.36                    & -0.02              & -0.07             & 0.28              & 0.12             & 0.21                         & -0.08             & \B-0.37             \\
$R^2$      & -0.23           & 0.40                  & \B0.60                  & -0.29                   & -0.37                   & 0.13               & 0.14              & -0.24             & 0.19             & -0.25                        & -0.08             & \B0.61              \\
\bottomrule 
 \label{tab:Corr-Show-a4-Eval-APRBSEqual15} \end{tabular}\end{table*}

\begin{figure*}[htbp]
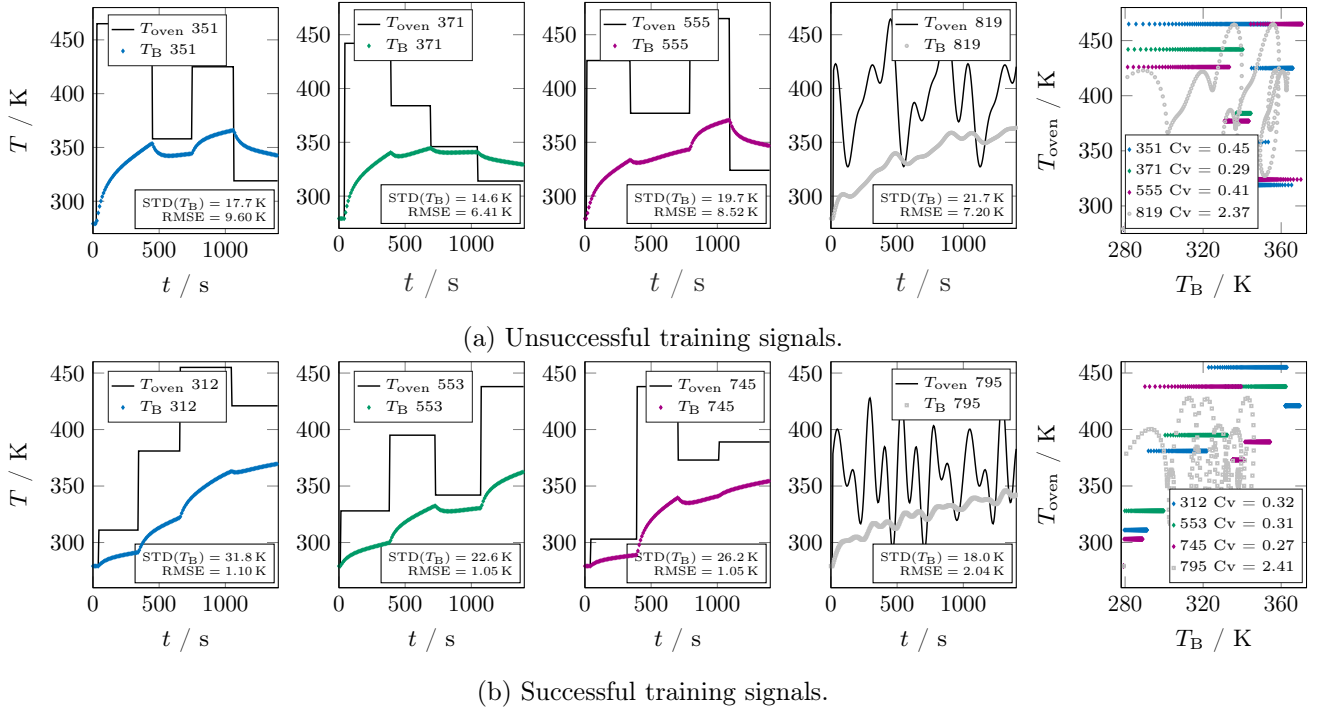

 \centering
  \begin{subfigure}[]{\textwidth}
  \setlength\fheight{0.20\textheight}
\setlength\fwidth{0.23\textwidth}
  \centering
\input{tiks/ROM-signal/351-loose.tex}
\input{tiks/ROM-signal/371-medium-good.tex}
\input{tiks/ROM-signal/555-loose.tex}
\input{tiks/ROM-signal/819-loose-MS}
\input{tiks/ROM-signal/looser-input.tex} 
       \caption{Unsuccessful training signals.}\label{fig:322-loose-a}
 \end{subfigure}
  \begin{subfigure}[]{\textwidth}
  \centering
\setlength\fwidth{0.23\textwidth}
\setlength\fheight{0.20\textheight}
\input{tiks/ROM-signal/312-win.tex}
\input{tiks/ROM-signal/553-win.tex}
\input{tiks/ROM-signal/745-win.tex}
\input{tiks/ROM-signal/795-win-MS}
\input{tiks/ROM-signal/winner-input.tex}
       \caption{Successful training signals.}\label{fig:322-loose-b}
 \end{subfigure}
 \caption{Illustration of successful and unsuccessful APRBS and multi-sine training cases and their respective input space coverage (right column). RMSE evaluated on test set AP15.}
 \label{fig:322-loose}
\end{figure*}

In the class of a priori KPIs, $\overbar{T_{\text{diff},j}}$ shows a strong correlation to error measures (compare~\autoref{fig:KPIandErrorCorrelations} (bottom-left)). It reveals that APRBS signals with a globally ascending trend guarantee lower ROM test errors. Remember that $\overbar{T_{\text{diff},j}}$ is the mean signed sum of delta jumps of an APRBS signal, where the first jump is excluded from the calculation. 
A similar interpretation seems valid when the best and worst training signals are analyzed over time and in their input space, as depicted in \autoref{fig:322-loose}. 
For example, signal 351 has high oven temperatures in the first steps and subsequently induces high surface temperatures at the run's end. Here, the model's evaporation non-linearity gets dominant. 
Good training signals seem to cover the top area above the main diagonal, and underperforming training signals the area above the secondary diagonal of the input space diagram (right column of \autoref{fig:322-loose}).

 \subsection{Signal type comparison} \label{sec:signaltype}
 
\begin{figure}[htbp]
 \centering
\setlength\fheight{0.25\textheight}
\setlength\fwidth{0.65\columnwidth}
 \input{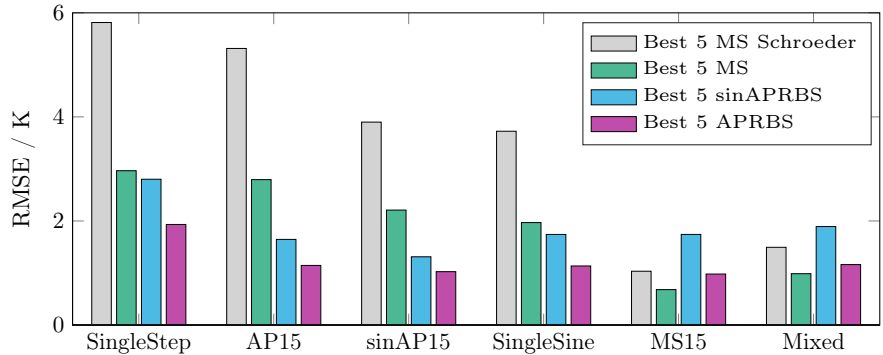}
  \caption{Mean RMSE of the best five training signals of their kind. Testing is performed on six different test sets. APRBS signals perform best from a global perspective.}
 \label{fig:best5bars}
\end{figure}

Error measures for the five best training signals per signal type were averaged. This renders the decision on what could be the most suitable signal type more robust for outliers. \autoref{fig:best5bars} shows the mean test RMSE for all available test sets. 
Obviously, step signals should not be used for non-linear system identification, which agrees with the literature~(\cite{Nelles2020}).
As expected, multi-sine trained ROMs perform best on multi-sine test sets. However, APRBS signals perform best when we consider all test scenarios. The best five APRBS signals have also been highlighted in red in \autoref{fig:KPIandErrorCorrelations} for better identification.
Based on these findings, the focus of the residual study shifts to multi-step signals, as they seem the most suitable signal type.
%

\subsection{sinAPRBS generated from APRBS degrade the test error}\label{sec:sinAPRBS}

\begin{figure}[htbp]
\centering
\setlength\fheight{0.25\textheight}
\setlength\fwidth{0.65\columnwidth}
 \input{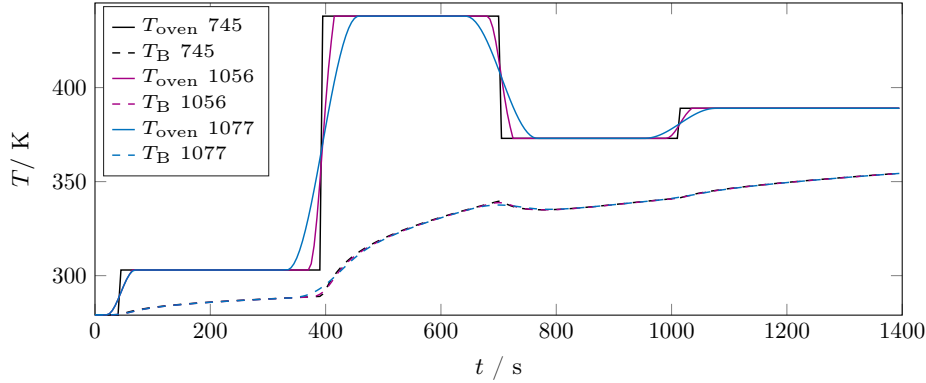} 
    \captionof{figure}{Transformation of APRBS signal 745 to a fast and slow sinAPRBS signal.}
    \label{fig:boxrms-745-sinstudy-a}

\end{figure}

In \autoref{fig:best5bars}, the best five sinAPRBS tend to perform worse than APRBS, even on sinAP15. This tendency is further investigated with the following test. The best 15 APRBS signals have been transformed into fast and slower sinAPRBS. \autoref{fig:boxrms-745-sinstudy-a} illustrates the transformation for APRBS 745. Observe the marginal differences in $T_\text{B}$. Nonetheless, 
\autoref{tab:sinAPRBSstudy} confirms the trend we have already discovered in the above sections: 
sinAPRBS signals increase the test RMSE with decreasing speeds of the sinusoidal transitions. This can be explained since rapid signals, such as steps, are considered to excite the broadest frequency range of a system. This property is taken from APRBS when sinusoidal transitions are inserted.

  \begin{table}[htbp]
\scriptsize
\centering
 \caption{Mean error measures of 15 APRBS ROMs and their transformed fast and slow sinAPRBS ROMs.} 
 \begin{tabular}{llll} \toprule 
& APRBS & sinAPRBS (fast) & sinAPRBS (slow) \\ 
\midrule 
RMSE / K & 1.63 & 2.32 & 2.90 \\ 
MAPE / - & 0.38 & 0.58 & 0.67 \\ 
MAX / K & 3.91 & 4.79 & 6.22 \\ 
MEDIAN / K & 0.15 & 0.48 & 0.14 \\ 
IQR / K & 1.64 & 2.46 & 3.36 \\ 
$\text{R}^2$ / - & 0.99 & 0.97 & 0.94 \\ 
\bottomrule 
 \label{tab:sinAPRBSstudy} \end{tabular}
\end{table}

\begin{figure}[htbp] 
 \centering
\setlength\fheight{0.25\textheight}
 \setlength\fwidth{0.65\columnwidth}
\input{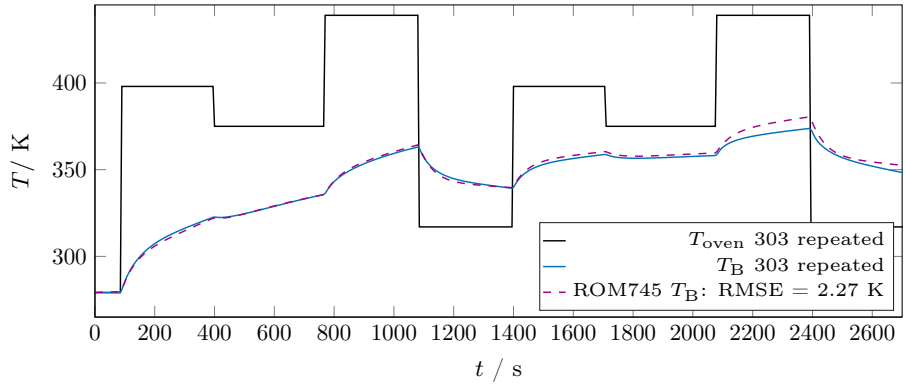}
  \caption{Extrapolation capability of the best 1-signal.}
 \label{fig:303-long}
\end{figure}

\subsection{DynROM investigations}\label{sec:romevals}
\paragraph{Extrapolation capability}
Extrapolation capability is a desired property for ROMs. The prediction remains valid, even if the trained parameter space is left. For a transient ROM, the trained time frame is also such a parameter. A simple test design is set up. One test signal  303 is catenated with a repetition of its own. \autoref{fig:303-long} depicts how the surface temperatures evolve for 1-signal-ROM745. 
One can observe the non-linear behavior of the physical model due to evaporation effects. In the second segment, the core and surface behave substantially differently from the first, although the excitation signal was repeated. The non-linear ROMs can extrapolate this behavior reasonably well, with RMSEs in the value range of the physical validation of the model (\autoref{fig:RabelerCub-b}).

A build-up of the prediction error from the onset of leaving the trained time window of \SI{1400}{s} can be observed. 
A possible source for the accumulating error is the extended non-linear behavior of our physical model at high temperatures. The surface evaporation acts as a heat sink and holds temperatures within the boiling temperature range.  Once the surface dries out, surface temperatures can heat up further.
If this behavior is intended to be included in the ROM more, the required time frame and higher oven temperatures would have to be included in the training.
\paragraph{Speed-up}

Preliminary results of our model predictive control architecture (results not shown) reveal that ROMs are required to solve within tenths of a second to predict a processing time of \SI{1800}{s}. For this simple cuboid model (with conjugate CFD not yet included), the requirement equals a necessary simulation speed-up of $\operatorname{Sp} \approx 10^{4}$. Speed-ups are defined in relation to physical process times, not the simulation times of the full-order model.
Previous investigations of conjugate open pan-frying in a realistic 3D setup required solving times of approximately two days on a 16-core cluster PC (compare \autoref{tab:predictionperformance}). 
The solution of the quarter cube model of this study takes approximately \SI{730}{s} to be solved on four cores of a modern laptop at full load. Conversely, the resulting FMU file provides simulations typically within \SI{100}{ms} ($\operatorname{Sp} \approx \SI{1.8E4}{}$) with no noticeable single-core computational load.
\newcommand{\specialcell}[2][l]{%
\begin{tabular}[#1]{@{}l@{}}#2\end{tabular}}

\begin{table}[hbtp]
\centering
\scriptsize
\addtolength{\tabcolsep}{-3.1pt}
\caption{Comparison of simulation and speed-up for a CFD conjugate heat transfer simulation (\cite{diss_simconf19}) of open-pan frying with a Finite Element and a DynROM simulation of the present challenge.}
\begin{tabular}{@{}llrlrrr@{}} \toprule
 Type        & Real time & Time &   Computational  load  & Problem info & $\mathrm{Sp}$\\ %
 \midrule
CFD       & \SI{2\,000}{s}  & \SI{181 212}{s} & \specialcell{parallel, full load \\ 16 $\times$ 3.2 GHz} & \si{588\,959} DOFs  & 0.01 \\ 
FEM 		& \SI{1800}{s} & \SI{730}{s} & \specialcell{parallel, full load \\ 4 $\times$ 3.2 GHz }  & \si{51\ 094} DOFs & 2.47 \\ %
 DynROM  & \SI{1800}{s}  & $\approx$ \SI{0.1}{s} & \specialcell{serial, not noticable\\ 1 $\times$ 3.2 GHz } &  FMU   &  $ \approx \SI{1.8E4}{}$\\ 
\bottomrule  
 \label{tab:predictionperformance}      
\end{tabular}
\end{table}

\begin{figure*}[htbp]
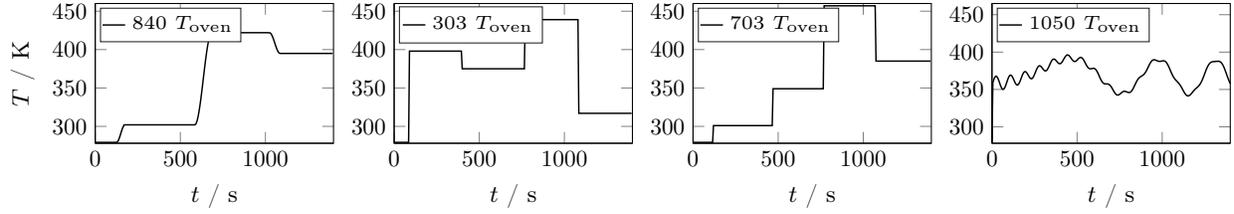
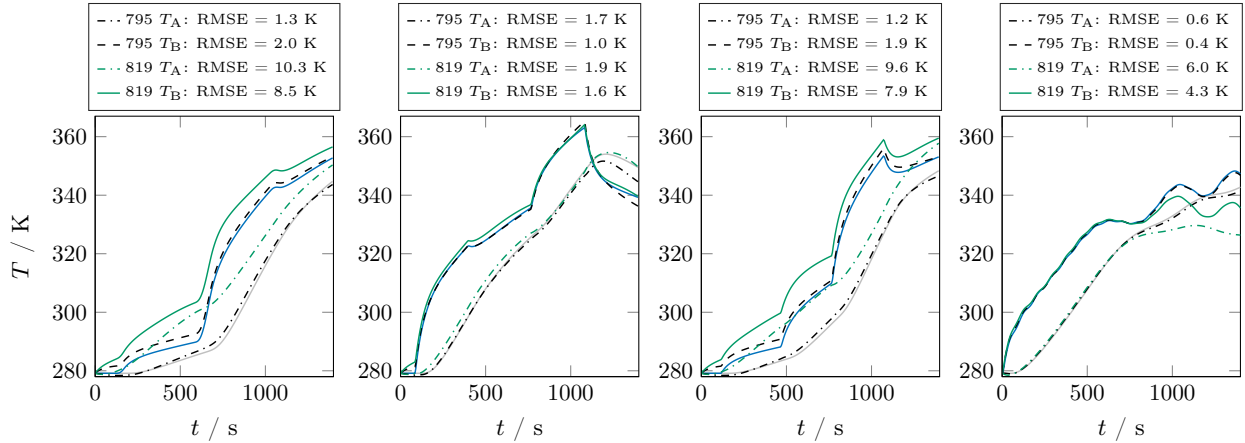
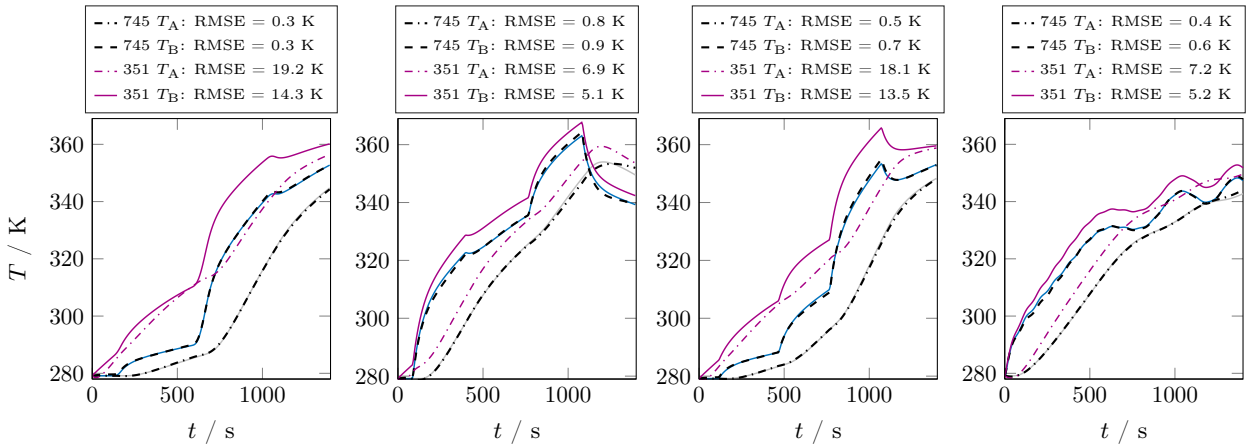
 
 \centering
   \begin{subfigure}[]{\textwidth}
  \centering
\setlength\fheight{0.15\textheight}
 \setlength\fwidth{0.27\textwidth}
      \input{tiks/ROM-signal/TimeEval-840sines2-Kopie}
       \input{tiks/ROM-signal/TimeEval-303sines-Kopie}
         \input{tiks/ROM-signal/TimeEval-703sines-Kopie}
     \input{tiks/ROM-signal/TimeEval-1050sines-Kopie}
       \caption{Test signals.}\label{fig:timeevals-a}
 \end{subfigure}
 
\vspace{0.5em}
   \begin{subfigure}[]{\textwidth}
  \centering
 \setlength\fheight{0.22\textheight}
 \setlength\fwidth{0.27\textwidth} 
     \input{tiks/ROM-signal/TimeEval-840sines2}
          \input{tiks/ROM-signal/TimeEval-303sines}
         \input{tiks/ROM-signal/TimeEval-703sines}
     \input{tiks/ROM-signal/TimeEval-1050sines}
       \caption{Testing of best (795) and worst (819) multi-sine 1-signal-ROM.}\label{fig:timeevals-b}
 \end{subfigure}
 
\vspace{0.5em}
     \begin{subfigure}[]{\textwidth}
  \centering
 \setlength\fheight{0.22\textheight}
 \setlength\fwidth{0.27\textwidth} 
     \input{tiks/ROM-signal/TimeEval-840AP}
       \input{tiks/ROM-signal/TimeEval-303AP}
      \input{tiks/ROM-signal/TimeEval-703AP}
     \input{tiks/ROM-signal/TimeEval-1050AP}
       \caption{Testing of best (745) and worst (351) APRBS 1-signal-ROM.}\label{fig:timeevals-c}
 \end{subfigure}

  \caption{Time evaluation of representative ROMs: selected test signals (a), multi-sine ROMs (b), step and multi-step ROMs (c). The full-order model solutions of $T_\text{A}$ and $T_\text{B}$ are represented by the grey and blue solid lines.}
 \label{fig:timeevals}
\end{figure*}

\paragraph{ROM evaluations over time}
To conclude the investigations, \autoref{fig:timeevals} depicts the evaluation of the best multi-sine and APRBS 1-signal-ROMs. 
Representative oven temperatures (first row) of sinAPRBS (left column), APRBS (central columns), and multi-sine (right column) signal types are chosen for testing.
The best multi-sine 1-signal-ROM795 evaluations for core (black, dash-dotted) and surface (black, dashed) temperatures can be found in \autoref{fig:timeevals-b}. \autoref{fig:timeevals-c} illustrates the evaluation of the best APRBS 1-signal-ROM745 (black dash-dotted and dashed lines). It can follow the core (grey) and surface (blue) temperatures of the full-order model with  $\text{RMSE} < \SI{1}{K}$. To elucidate once more the importance of proper signal selection for ROM training, ROM evaluations for the worst APRBS (red) and multi-sine (green) ROM were also displayed in \autoref{fig:timeevals}.

\section{Conclusion and outlook}\label{sec:conclusionandoutlook}

\paragraph{Results and discussion}
The presented method can provide physics-based data-driven ROMs with speed-ups of $\operatorname{Sp} \approx \SI{1.8E4}{}$ compared to real-time simulation. With characteristic solution times of one-tenth of a second, the faster-than-real-time simulations enable the ROM usage in \enquote{on-device} model predictive control. The computational load is not noticeable on a single-core processor.
Meanwhile, $\text{MAPE} = \SI{0.2}{\%}$, and RMSEs less than \SI{1}{K} can be reached. The accuracy is better than the precision of standard industrial thermocouples, and the test RMSE is significantly less than the validation error of the utilized food model.

When considering a suitable excitation signal type for data-driven ROM generation, one must distinguish ROM training, testing, and operation.
APRBS signals should be used for training. They show the lowest mean test errors on APRBS, sinAPRBS, single-step or single-sine test sets.
Testing should be performed on unbiased, fair test data sets. Uniform distribution of output variable can be obtained with a $\chi^2$-based signal picking. 
sinAPRBS emulate the operational temperatures in convection ovens best. APRBS-trained ROMs show even lower test errors on sinAPRBS than on APRBS test sets.
This finding contradicts common recommendations in the literature, where signal types close to operational signals are suggested for training (\cite{sig_Gringard2016}).
Multi-sines may be appropriate training signals if the operational scenario is purely sinusoidal.

This article focuses on one cornerstone of our proposed Digital Twin framework: parsimonious design of experiments that ensure a high-fidelity ROM generation. 
Compared to many other machine learning approaches, this method requires substantially less training data: only one signal of \SI{1400}{s} in length. 
A key feature of the procedure is the a priori analysis of the full-order model training data. KPIs support the selection of the most suitable training signal.
A common recommendation in literature is the even coverage of input or output space of a model (\cite{Nelles2020,sig_Talis2021}). We demonstrated that this paradigm might not hold for the utilized thermal processing model.
Alternatively, high standard deviation in surface temperatures or the signed sum of the oven temperatures in APRBS signals can serve as good KPIs to select training signals. The two KPIs correlate best with a ROM's test error, $R=-0.76$ and $R=-0.68$, respectively.

This study potentially serves as a first guidance for ROM training design of experiments in the context of convection oven heating processes.
Similar trends of correlations are expected for similar heating scenarios, as long as heat input and loss mechanisms, effective transport properties, porosity, moisture content, operational temperatures and time remain in comparable value ranges. 
If, for example, a different product should be processed autonomously in an existing smart cooking device, one may want to exchange the simulation model. 
If so, the presented framework is capable of doing a complete model exchange. We suggest the following course of action:
\begin{outline}
\1 Consider the maximum total process time and divide it into a reasonable amount of input steps / operational conditions (such as four steps in this study, taking into consideration the largest time constant of the model).
\1 APRBS signals should be the preferred training signals for non-linear system identification unless testing and application are exclusively sinusoidal.
\1 Consider sinAPRBS as an appropriate test signal if the physical process device cannot realize fast steps of the input signal (such as the thermal lag of the oven temperature in the presented case).
\1 A significant number of pseudo-random signals should be generated (such as 50 - 100 in this study) to excite the full-order model.
\1 Generate a 1-signal-ROM for each signal and test the ROM upon the complementary, residual data sets. 
\1 In the case of different physics compared to the presented convection oven scenario, a novel correlation between signal property and test RMSE may have to be found.
\1 A fair test subset of signals may be curated, as outlined in \autoref{sec:trainandeval}.
\1 The signals should be sorted based on their KPIs, such as high STD($T_\text{B}$)-signals being listed first.
\1 Select the best performing 1-signal-ROM as the base case and combine it with other one or two high KPI signals to further reduce the test error.
\1 Repeat combinatory rotation of signals until test error reduction stagnates.
\end{outline}

\paragraph{Contributions}
Overall, the presented Digital Twin framework for thermal processing contributes to shifting the perspective on Digital Twins towards the autonomy of processes. 
The method is designed to be applicable in the industry and is universal to modeling software. This is especially beneficial for food science and technology, where various numerical approaches and software are used to model food processes (\cite{pm_datta_toward2016}).
Our code framework \enquote{TwinLab} automates signal synthesis, ROM training and testing. Interfaces for full-order models in COMSOL Multiphysics or ANSYS Fluent and subsequent ROM generation with ANSYS DynROM are implemented. The source code will be made publicly available on GitHub.
The resulting ROMs are stored in the encrypted FMU format. The small files can be easily deployed and updated via software updates of the processing devices.
From a more global perspective, the method enables Digital Twin-based autonomy to a much broader spectrum of processes, not restricted to food science.

\paragraph{Outlook}\label{sec:outlook}

Four future directions will be taken to approach the completion of the presented framework:
\begin{itemize}
\item On the foundation of the developed KPIs, the combination of training signals may be investigated to further reduce the ROM test error.
\item The extension to realistic shapes seems promising. The three-dimensional full-order model data can be compressed with proper orthogonal decomposition. The time-dependent mode shape coefficients may be trained in ROMs in the same manner as the temperatures in this study.

\item The presented ROM speed-ups allow deriving an intelligent decision-making algorithm based on model predictive control that is executable on device level.

\item The utilized thermal processing model by \cite{cm_rabeler_mod2018} lumps convection and radiation mechanisms in a heat transfer coefficient. The heating mechanisms cannot be controlled separately. 
This is a common practice observed within food science literature. We believe that conjugate CFD models -- hence, coupling food processing models with explicitly simulated thermal CFD -- will substantially increase model quality. Ultimately, a more realistic model enables enhanced Digital Twin-based control of the process.

\end{itemize}

\section*{Author contributions}
Conceptualization, M.K.; methodology, M.K..; software, M.K., M.P.; validation, M.K.; formal analysis, M.K.; investigation, M.K.; data curation, M.K., M.P.; writing -- original draft preparation, M.K.; writing -- review and editing, M.K.; visualization, M.K.; supervision of M.P., M.K.; supervision, M.S.; All authors have read and agreed to the published version of the manuscript.

\section*{Acknowledgments}

The work of Maximilian Kannapinn is supported by the Graduate School CE within the Centre for Computational Engineering at Technische Universität Darmstadt. Maximilian Kannapinn would like to thank Christoph Petre and Valery Morgenthaler for their information on Dynamic ROM Builder.

\section*{Declarations of interest} none

\appendix

\section{Swelling pressure model for thermal processing of chicken fillet}\label{sec:chickenmodel}

The conservation of water concentration $c$ (\si{mol.m^{-3}}) and energy, here for temperature $T$ (\si{K}), reads:
\begin{align}
\label{eq:conservation-differential-appendix}
\frac{\partial c}{\partial t} + \frac{\partial}{\partial x_i} \left(c  u_i  \right) &= \frac{\partial}{\partial x_i} \left(D_\text{cb} \frac{\partial c}{\partial x_i} \right)\,,\\
\label{eq:energy-appendix}
\rho c_\text{p}  \frac{\partial T}{\partial t} + \rho_\text w c_\text{p,w} u_i \frac{\partial T}{\partial x_i}  &=  \frac{\partial}{\partial x_i} \left(\lambda_{ij} \, \frac{\partial T}{\partial x_j}\right)\,.
\end{align}
Specific heat $c_{\text{p},i}$ and thermal conductivity $\lambda_{ij}$ of the food components protein, ash, fat and water are modeled temperature-dependent with formulas by \cite{cm_choiokos1986}. The fiber orientation of the fillet (horizontal direction of \autoref{fig:RabelerCub-a}) is accounted for with the parallel and serial model when calculating the effective thermal conductivity,
\begin{linenomath*}
\begin{align}
    \lambda_\text{parallel} =  \sum_{i=1}^n \phi_i\lambda_i  ,\qquad
    \frac{1}{\lambda_\text{perp.}} =  \sum_{i=1}^n \frac{\phi_i}{\lambda_i}\,,
\end{align}
\end{linenomath*}
where $\phi_i$ is the respective volume fraction.
The effective specific heat is obtained from
\begin{align}
c_\text{p} =  \sum_{i=1}^n y_i\, c_{\text{p},i}\,,
\end{align}
where $y_i$ is the respective mass fraction of the component.
The mass fractions of protein (0.21), ash (0.01), and fat (0.01) remain fixed, whereas their volume fractions vary dynamically with varying moisture concentrations.
\begin{table*}[htbp]
\scriptsize
\centering
 \caption{Constants and initial conditions}
 \begin{tabular}{lllll} \toprule 
Variable& Name & Value & Constant & Value\\ 
\midrule 
$C_0$ 		& Initial moisture mass fraction & $ C_0 = 0.76 \Rightarrow c_0=\frac{C_0\, \rho_\text{cb}} { M_w}$ 			& $T_\sigma $  		& \SI{315}{ K}\  \\ 
$T_0$ 		& Initial temperature & $\SI{279.15}{K}$ 				& $\overline{T}  $  		& \SI{342.15 }{ K}  \\ 
$\rho$ 		& Effective density & \SI{1050}{kg.m^{-3}}						& $\Delta T  $  		& \SI{4}{ K}\  \\ 	
$D_\text{cb}$ 	& Effective Diffusity & $\SI{3e-10}{m^2.s^{-1}}$ 				 & $G^\prime_\text{max} $  		& \SI{92000}{Pa} \\ 
$h_\text{amb}$ & Heat transfer coefficient & $ 44\; (50)\, \si{W.m^{-2}.K^{-1}}$ (bottom)	& $G^\prime_\text{0} $  		& \SI{13500 }{ Pa} \\ 
$H_\text{evap}$ &  Latent heat of evaporation & $ \SI{2.3E6}{J/kg}$ 	& $\kappa  $  		& 	$\SI{3E-17}{m^2} $ \\ 
$a_w$  & Water activity & $ 1 - \frac{0.073}{M_\text{db}} $ 	& $\mu_w$  &	$\SI{0.988E-3}{kg.m^{-1}.s^{-1}} $	   \\
 & & & $\rho_w$ & $\SI{998}{kg.m^{-3}} $ \\
\bottomrule 
 \label{tab:ctconstants} \end{tabular}\end{table*}
The boundary conditions for mass and energy conservation read
\begin{linenomath*}
\begin{align}
   \left( c u_i - D_\text{cb} \frac{\partial c}{\partial x_i} \right) N_i = -\frac{m_\text{evap}} {M_\text{w}} = -\frac{\beta_\text{tot}}{ M_\text{w} } \left( x_\text{v,surf} - x_\text{\text{v,amb}} \right)  \\
   = -\beta_\text{tot} \left( a_w \, \frac{p_\text{sat}(T_{\text{surf}}) }{R\,T_{\text{surf}} } - \Phi \, \frac{p_\text{amb} }{R\,T_\text{oven}} \right)\,, \\
   \left(\rho c_\text{p}  \,u_i -\lambda_{ij} \, \frac{\partial T}{\partial x_j}\right) N_i =  h_\text{amb}(T_\text{oven}-T) - m_\text{evap} H_\text{evap}\,,
\end{align}
\end{linenomath*}
where $p_\text{sat}$ and $p_\text{amb}$ are the vapor pressures at the surface and in the oven. $R$ is the universal gas constant. 
Oven conditions have a fixed relative humidity of $\Phi=0.05$

The total mass transfer coefficient is determined via the serial model.
The external mass transfer coefficient is determined from the Chilton-Colburn relation:
\begin{linenomath*}
\begin{align}
  \beta_\text{ext} = \frac{h_\text{amb}}{\rho_\text{amb} c_\text{p,amb}} \Lewi ^{-2/3},  \quad
  \Lewi &= \frac{\Schm}{\Pran} \approx 0.91
\end{align}
\end{linenomath*}
where $\Lewi$, $\Schm$ and $\Pran$ are the Lewis, Schmidt, and Prandtl number.
The skin coefficient
\begin{linenomath*}
\begin{align}
   \beta_\text{skin} &= 0.04\, C^{F(T)}\,, 
  \end{align}
  \end{linenomath*}
 accounts for surface resistance to evaporation with a function $F(T)$ which transitions the exponent from $7$ to $1$ when the boiling temperature is reached.  
The serial model for the total mass transfer coefficient 
\begin{linenomath*}
\begin{align}
  \beta_\text{tot} = \frac{1}{\frac{1}{\beta_\text{ext}} + \frac{1}{\beta_\text{skin}}}
\end{align}
\end{linenomath*}
pronounces $\beta_\text{skin}$ for small C (dry surfaces) and reduces the total mass transfer coefficient to a value considerably less than $\beta_\text{ext}$. 

\paragraph{Swelling pressure flow in porous media}

Darcy's law is used to relate the velocity $u_i$ to a pressure gradient $\nabla p$: 
\begin{linenomath*}
\begin{align}\label{eq:ui}
u_i &= -\frac{\kappa}{\mu_w}\,\frac{\partial p}{\partial x_i} \,,
\end{align}
\end{linenomath*}
where $\kappa$ denotes the permeability of the matrix and $\mu_w$ is the viscosity of water.
Protein denaturation induces shrinkage during a roasting process, which results in water exudation at the surface.  
For thermal processing applications, the internal pressure  $p$ due to shrinkage can be related to the water concentration in meat 
as
\begin{linenomath*}
\begin{align}\label{eq:p}
p &= G^\prime(T) \left( C - C_{eq}(T)\right) \,,\\
C_\text{eq}(T) &= C_{0} - \frac{0.31}{1+ 30\, \text{exp}\left(-0.17 \left(T-T_\sigma\right)\right) }\,,\\
G^\prime(T)&= G^\prime_\text{max} + \frac{G^\prime_0 - G^\prime_\text{max}}{1 + \text{exp} \left(\frac{T-\overline{T}}{\Delta T}\right) }\,,
\end{align}
\end{linenomath*}
where $C_\text{eq}$ is water holding capacity and $G^\prime(T)$ is a modeled storage modulus.
The required constants are summarized in \autoref{tab:ctconstants}. 
For more details on the food model, the reader is referred to \cite{cm_rabeler_mod2018}.




\renewcommand*{\bibfont}{\footnotesize}
\printbibliography

%
%
%
%
\end{document}